\documentclass[a4paper,fleqn,usenatbib]{mnras}

\usepackage[normalem]{ulem}

\usepackage{newtxtext,newtxmath}
\usepackage{siunitx}
\usepackage[T1]{fontenc}
\usepackage{ae,aecompl}

\usepackage{graphicx}	
\usepackage{amsmath}	
\usepackage{pdflscape}
\usepackage{hyperref}

\defcitealias{2019MNRAS.490.3806N}{Paper\,I}
\defcitealias{2020MNRAS.495.4924N}{Paper\,II}

\title[PATHOS\,III: Young Associations]{A PSF-based Approach to TESS High
  quality data Of Stellar clusters (PATHOS) - III. Exploring the
  properties of young associations through their variables, \textit{dippers}, and candidate exoplanets.}

\author[D.\ Nardiello]{D.\ Nardiello$^{1,2}$\thanks{E-mail: domenico.nardiello@lam.fr} \\
$^{1}$Aix Marseille Univ, CNRS, CNES, LAM, Marseille, France \\
$^{2}$Istituto Nazionale di Astrofisica - Osservatorio Astronomico di Padova, Vicolo dell'Osservatorio 5, IT-35122, Padova, Italy \\
}

\date{Accepted 2020 September 2. Received 2020 September 2; in original form 2020 July 31}

\pubyear{2020}
\begin{document}
\label{firstpage}
\pagerange{\pageref{firstpage}--\pageref{lastpage}}
\maketitle

\begin{abstract}
  Young associations in star forming regions are stellar systems that
  allow us to understand the mechanisms that characterise the stars in
  their early life and what happens around them. In particular, the
  analysis of the disks and of the exoplanets around young stars
  allows us to know the key processes that prevail in their
  evolution and understand the properties of the exoplanets orbiting
  older stars. The {\it TESS} mission is giving us the opportunity to
  extract and analyse the light curves of association members with
  high accuracy, but the crowding that affects these regions
  makes difficult the light curve extraction. In the PATHOS project,
  cutting-edge tools are used to extract high-precision light curves
  and identify variable stars and transiting exoplanets in open
  clusters and associations. In this work, I analysed the light curves
  of stars in five young ($\lesssim 10$~Myr) associations, searching
  for variables and candidate exoplanets. By using the rotational
  periods of the association members, I constrained the ages of the
  five stellar systems ($\sim 2$-$10$~Myr). I searched for {\it
    dippers} and I investigated the properties of the dust that forms
  the circumstellar disks. Finally, I searched
  for transiting signals, finding 6 strong candidate exoplanets. No
  candidates with radius $R_{\rm P}\lesssim 0.9~R_J$ have been
  detected, in agreement with the expectations. The frequency of giant
  planets resulted to be $\sim$2-3~\%, higher than that expected for
  field stars ($\lesssim 1$~\%); the low statistic makes this
  conclusion not strong, and new investigations on young objects are
  mandatory to confirm this result.
 \end{abstract}



\begin{table*}
  \caption{Association information}
  \resizebox{0.99\textwidth}{!}{
    \begin{tabular}{l c c r r r r r r}
  \hline
  Association Name & $\alpha_0$ & $\delta_0$ & $r_0$ & $\mu_\alpha \cos{\delta} $ & $\mu_{\delta}$ & $\pi$ & [Fe/H]$^{\rm (1)}$ & $N$ \\
  &  (deg.)    &  (deg.)   &  (deg.) &  (mas\,yr$^{-1}$)        &  (mas\,yr$^{-1}$) & (mas)  & &   \\
\hline
ChI     &     166.70      &  $-77.30$  &  7.0  & $-22.4\pm 1.4$ & $+0.5 \pm 2.1$ & $5.22 \pm 0.15$ & $-0.08 \pm 0.04$ & 191 \\
ChII    &     193.41      &  $-77.17$  &  7.0  & $-20.2\pm 2.8$ & $-7.5 \pm 2.7$ & $5.04 \pm 0.13$ & $-0.11 \pm 0.14$ & 50 \\
Lup     &     240.00      &  $-38.25$  & 25.0  & $-11.5\pm 0.6$ & $-21.8\pm 0.6$ & $5.42 \pm 1.24$ & $-0.10 \pm 0.04$ & 3105 \\ 
Vel     &     122.34      &  $-47.35$  & 10.0  & $-5.4 \pm 0.9$ & $+8.7 \pm 0.9$ & $2.53 \pm 0.31$ & $-0.06 \pm 0.02$ & 2895 \\
CrA     &     285.46      &  $-36.98$  & 15.0  & $+1.2 \pm 3.3$ & $-27.2\pm 1.6$ & $6.61 \pm 0.28$ & $-0.04 \pm 0.05$ &  388 \\
\hline
\multicolumn{9}{l}{$^{\rm (1)}$~Metallicities from James et al.~(2006), Spina et al.~(2014a,b)} \\
\end{tabular}

  }
  \label{tab:1}
\end{table*}

\section{Introduction}

To date, more than 4000 exoplanets have been discovered and
characterised\footnote{\url{https://exoplanetarchive.ipac.caltech.edu/}},
but their properties have not always been those we observe
today. Indeed, the exoplanets we observe were born with different
properties: in their early life, planets are subject to a series
of interactions with other bodies or the host star, that cause
changing in their orbital and physical parameters (migration,
planetary impacts, atmospheric photoevaporation, etc.). All these
processes have been studied in details (see, e.g.,
\citealt{2007ApJ...654.1110T,2010ApJ...719..810I,2012ApJ...751..158H,2013ApJ...776....2L,2013ApJ...775..105O,2015Icar..247...81S,2018haex.bookE.141S}
) and partially explain some observables, like, e.g., the gap in the
radius distribution of small planets at $1.5$-$2.0~R_{\earth}$
(\citealt{2017AJ....154..109F,2018AJ....156..264F}), the dearth of
short-period giant planets in close-in exoplanet distribution (see,
e.g., \citealt{2018MNRAS.479.5012O} and references therein), and the
accretion of the gaseous envelopes for giant planets
(\citealt{2003A&A...402..701B,2007ApJ...655..541M,2012ApJ...745..174S,2017A&A...608A..72M}).

In order to understand all the mechanisms that prevail in the life of
an exoplanet, it is mandatory to search for and monitor stars having
different ages. Unfortunately, stellar age is one of the most
difficult parameter to measure, unless the star is member of an
association or of a star cluster (open or globular): in the latter cases, the age of the
star can be well constrained thanks to the use of theoretical models.
For this reason, the interest on these objects has grown in recent
years and many photometric and spectroscopic works have been carried out on their members until
now (e.g., \citealt{2012ApJ...756L..33Q,2013Natur.499...55M,2014ApJ...787...27Q,2016AJ....151..112D,2016ApJ...818...46M,2016A&A...588A.118M,2016MNRAS.461.3399P,2018AJ....155....4M,2018AJ....155...10C,2018AJ....156...46V,2019A&A...630A..81B,2019ApJ...880L..17N,2020MNRAS.495..650G}).

The {\it Kepler} (\citealt{2010Sci...327..977B}) and {\it K2}
(\citealt{2014PASP..126..398H}) missions were a success, allowing the
detection of many exoplanets, also around stellar cluster and
association members
(\citealt{2013Natur.499...55M,2016A&A...594A.100B,2016AJ....152..223O,2016AJ....152...61M,2016MNRAS.463.1831N,2016MNRAS.463.1780L,2017AJ....153..177P,2018AJ....155..173C,2019AJ....158...79D,2019ApJ...885L..12D}), but
their sky coverage was limited. The {\it Transiting Exoplanet Survey
  Satellite} ({\it TESS}, \citealt{2015JATIS...1a4003R}) mission is
giving us the opportunity to study stellar cluster and association
members with high photometric accuracy and unprecedented sky and temporal
coverage: the satellite has probed more than 80~\% of the sky in its
first two years of mission, observing a large fraction of stellar
clusters and associations of the Galaxy for $\sim 27$ days or more,
and on July 2020 has started its extended mission. Given the
low resolution of the {\it TESS} images and the high-levels of star
crowding typical of clusters/associations, the extraction of high
precision light curves from {\it TESS} data needs appropriate
techniques, like the use of the difference imaging analysis
(\citealt{2019ApJS..245...13B}) or point spread function (PSF) models
(\citealt{2019MNRAS.490.3806N}).

The project 'A PSF-based Approach to {\it TESS} High Quality data Of
Stellar clusters’ (PATHOS; \citealt{2019MNRAS.490.3806N}, hereafter
Paper~I) is aimed at finding and characterisation of candidate
exoplanets and variable stars in stellar clusters and associations, by
using high-precision light curves obtained with a cutting-edge tool
based on the use of empirical PSFs and neighbour subtraction. This
technique allows us to minimise the dilution effects due to neighbour
contaminants, and extract high precision photometry even for faint
stars ($T\sim 17$-$18$). The efficiency of the method was demonstrated
in \citetalias{2019MNRAS.490.3806N}: high precision light curves of
stars located in an extreme crowded region centred on the globular
cluster 47\,Tuc, containing also Galactic and Small Magellanic Cloud
sources, were analysed. Many variables and one candidate hot-Jupiter
were identified. Using the same technique, \citet[hereafter
  Paper~II]{2020MNRAS.495.4924N} searched for exoplanets among the
light curves of $\sim 163\,000$ stellar members of 645 open clusters
observed during the first year of {\it TESS} mission, finding 11
strong candidates in eight open clusters with ages between $\sim
30$~Myr and $\sim 2$~Gyr.

In this third work of the series, I analysed the properties of the
members of five young ($\lesssim 10$~Myr) associations in as many star
forming regions by using the light curves extracted from the images
collected during the first year of the {\it TESS} mission. The
analysed associations are: Chamaeleon~I, Chamaeleon~II, Lupus, $\gamma$
Velorum, and Corona Australis associations. Given their young ages
($\lesssim 10$~Myr), these associations host a large number of T-Tauri
pre-main sequence stars, and for this reason they are also known as
{\it T-associations}, term coined by \citet{1949AZh....26....3A} in
his study on the importance of stellar associations for the
understanding of the stellar formation and evolution. Today, the study
of the properties of the young association members allows us not only
to investigate the life of the stars, but also how circumstellar disks
and exoplanets are born and evolved around them. Therefore, the
analysis of the {\it TESS} light curves of young stellar objects in
star forming regions offers the unique opportunity to trace the origin
and early evolution of circumstellar disks and exoplanets orbiting
them. In the last years, a large number of studies concentrate their
attention on young associations aimed to explore the metal content of
their stars (e.g.,
\citealt{2006A&A...446..971J,2008A&A...490.1135G,2008A&A...480..889S,2009A&A...501..973D,2011A&A...526A.103D,2011A&A...530A..19B,2012MNRAS.427.2905B,2012A&A...547A.104B,2014A&A...567A..55S,2014A&A...568A...2S,2017MNRAS.464.1456J}),
the disks that surround their young members (e.g.,
\citealt{2005ApJ...623..493C,2006ApJ...651L..49C,2009ApJ...705.1646C,2011ApJ...738..122C,2012ApJ...758...31L,2016ApJ...816...69A,2017MNRAS.470..202B,2018MNRAS.480.5099K,2019A&A...624A..87B,2020arXiv200703393A,
  2020MNRAS.tmp.1773B}), and the variability of the main sequence
stars (e.g.,
\citealt{2018AJ....155..196R,2019AJ....158...77C,2020AJ....159..273R})
in order to constraint their ages. Even if, in the last years,
associations and stellar clusters have been the subjects of many
exoplanet surveys, just a handful (candidate) exoplanets are known to
orbit members of young ($\lesssim 100$-150~Myr) clusters and
associations, and just one exoplanet orbits a (pre-)main sequence star
in a $\lesssim 10$~Myr old association, K2-33b (Upper Scorpius
association, \citealt{2016Natur.534..658D,2016AJ....152...61M}).
Other known exoplanets and candidates in young systems are: the hot
Jupiter HIP~67522 (Sco-Cen association, $\sim 17$~Myr,
\citealt{2020AJ....160...33R}), the planetary system around the star
V1298~Tau (Tau-Aur association, $\sim 23$\,Myr,
\citealt{2019ApJ...885L..12D}), the two candidate exoplanets PATHOS-30
and PATHOS-31 (IC~2602, $\sim 35$~Myr,
\citetalias{2020MNRAS.495.4924N}), the Neptune-size exoplanet
DS~Tuc~Ab (Tuc-Hor association $\sim 40$~Myr,
\citealt{2019ApJ...880L..17N, 2019A&A...630A..81B}), and the
sub-Neptune EPIC~247267267b (Cas-Tau group, $\sim 120$~Myr,
\citealt{2018AJ....156..302D}).

\begin{figure*}
  \includegraphics[bb=17 161 588 716, width=0.8\textwidth]{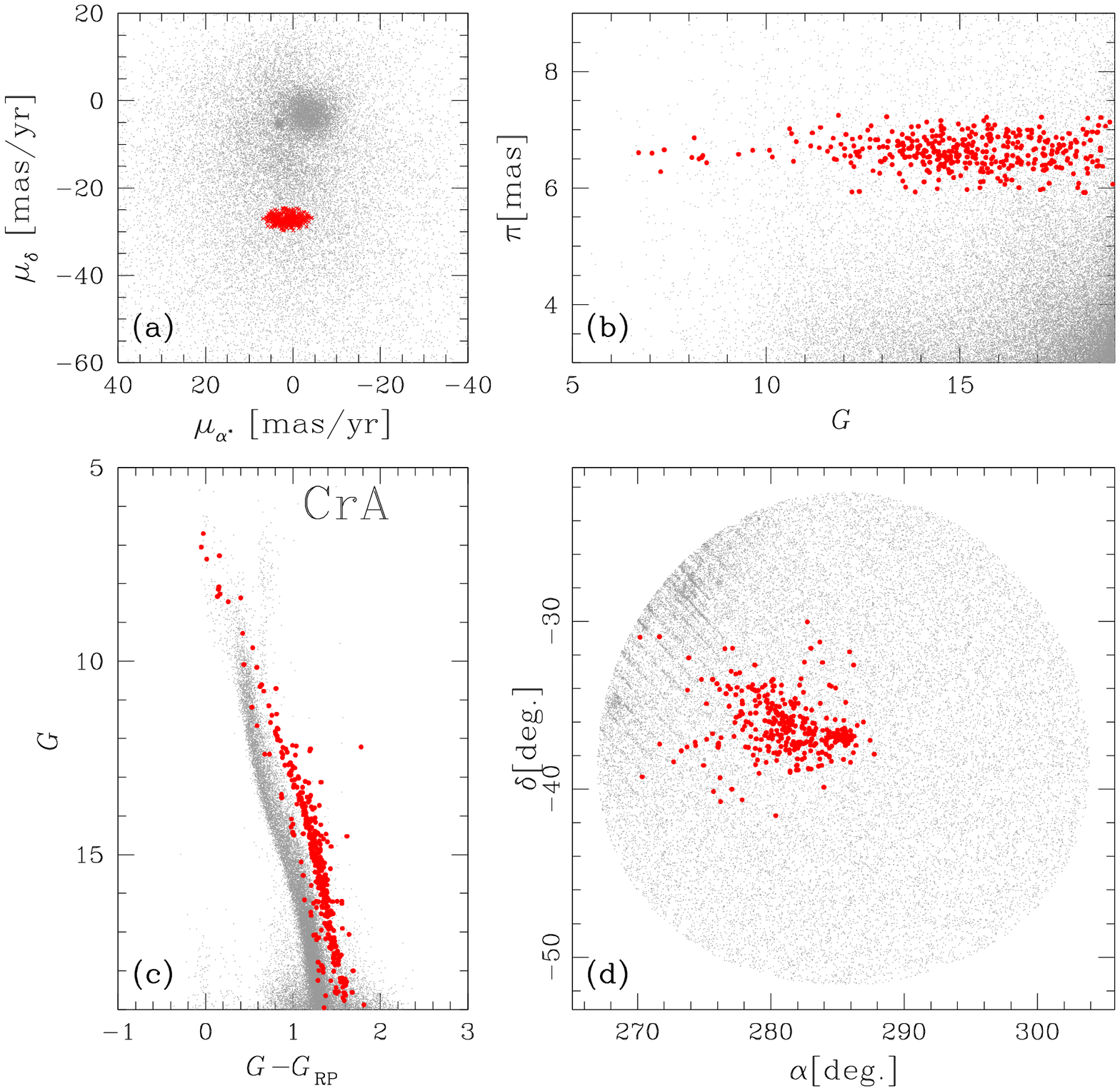}
  \caption{Overview on the selection procedure of likely CrA association
    members. Panel (a) shows the vector-points diagram of proper
    motions for the stars in the circular region ($r_0=15.0$ deg.)
    of the sky centred on $(\alpha_0, \delta_0)=(285.46,-36.98)$;
    panel (b) is the parallax distribution of the same stars as a
    function of the $G$ magnitude; panel (c) and (d) show the $G$
    versus $(G-G_{\rm RP})$ CMD and the $(\alpha, \delta)$ positions for the
    stars in the considered region. The red and the grey points
    represent the likely association members and the discarded stars
    in the selection procedure, respectively. In all the panel, for clarity, only
     20\% of the discarded stars are plotted. \label{fig:0}}
\end{figure*}

\begin{figure*}
  \includegraphics[bb=17 451 561 704, width=0.9\textwidth]{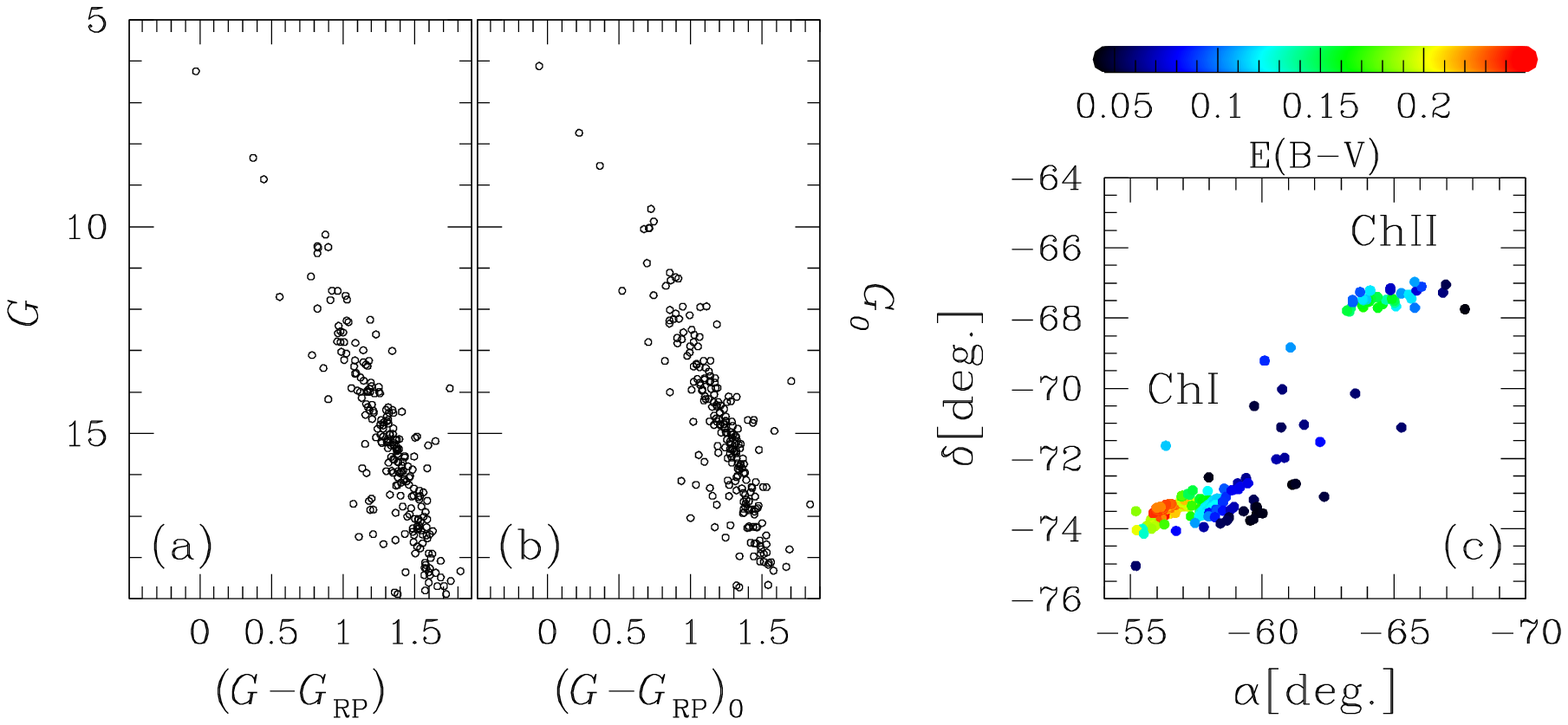}
  \caption{Reddening correction for the associations ChI and ChII
    based on the dust-map released by
    \citet{2016ApJ...818..130B}. Panels (a) and (b) show the $G$
    versus $(G-G_{\rm RP})$ CMD before (panel (a)) and after (panel
    (b)) the correction. Panel (c) shows the reddening map for the two
    associations.
    \label{fig:1}}
\end{figure*}

In the present work, the PATHOS pipeline is used to extract and correct the
{\it TESS} light curves of a sample of likely young association
members (Sect.~\ref{sec:obs}). Variable stars have been identified in
order to constraint the association ages and analyse the dust in
 the circumstellar disks around young stellar objects
(Sect.~\ref{sec:var}). The discovery and characterisation of new
candidate exoplanets orbiting stars in the aforementioned associations
and their frequency is reported in
Sect.~\ref{sec:cand_exo}. Section~\ref{sec:sum} is a summary and a
discussion of the results obtained in this work.

\section{Observations and data reduction}
\label{sec:obs}

In this work, I extracted the light curves of the stars in five very
young associations observed during the first year of the {\it TESS}
mission. In particular, I used Full Frame Images (FFIs) collected
during Sectors 7, 8, 9, 11, 12, 13. I produced a total of 7150 light
curves associated to 4459 stars.  The pipeline adopted for the light
curve extraction and correction is widely described in
\citetalias{2019MNRAS.490.3806N} and \citetalias{2020MNRAS.495.4924N}.
The pipeline includes the use of the light curve extractor
\texttt{IMG2LC}, developed by
\cite{2015MNRAS.447.3536N,2016MNRAS.455.2337N} for ground-based images
and improved by \citet{2016MNRAS.456.1137L,2016MNRAS.463.1780L} and
\citet{2016MNRAS.463.1831N} for {\it Kepler/K2} space-based
data. Briefly, the routine uses empirical Point Spread Functions
(PSFs) and an input catalogue (see Section~\ref{sec:input_cat}) to
extract aperture and PSF-fitting photometries of each star in the
catalogue after the subtraction of all the neighbours from each {\it
  TESS} FFI.  The raw light curves are then
corrected for systematic effects by fitting and applying to them the
Cotrending Basis Vectors described in
\citetalias{2020MNRAS.495.4924N}.  Light curves will be released in
\texttt{ascii} and \texttt{fits} format on the Mikulski Archive for
Space Telescopes (MAST) as a High Level Science Product (HLSP) under
the project
PATHOS\footnote{\url{https://archive.stsci.edu/hlsp/pathos}} (DOI:
10.17909/t9-es7m-vw14). A description of the format of the light
curves is reported in \citetalias{2019MNRAS.490.3806N} and
\citetalias{2020MNRAS.495.4924N} and in the MAST webpage of the PATHOS
project.

\begin{figure*}
  \includegraphics[bb=17 305 577 704, width=0.9\textwidth]{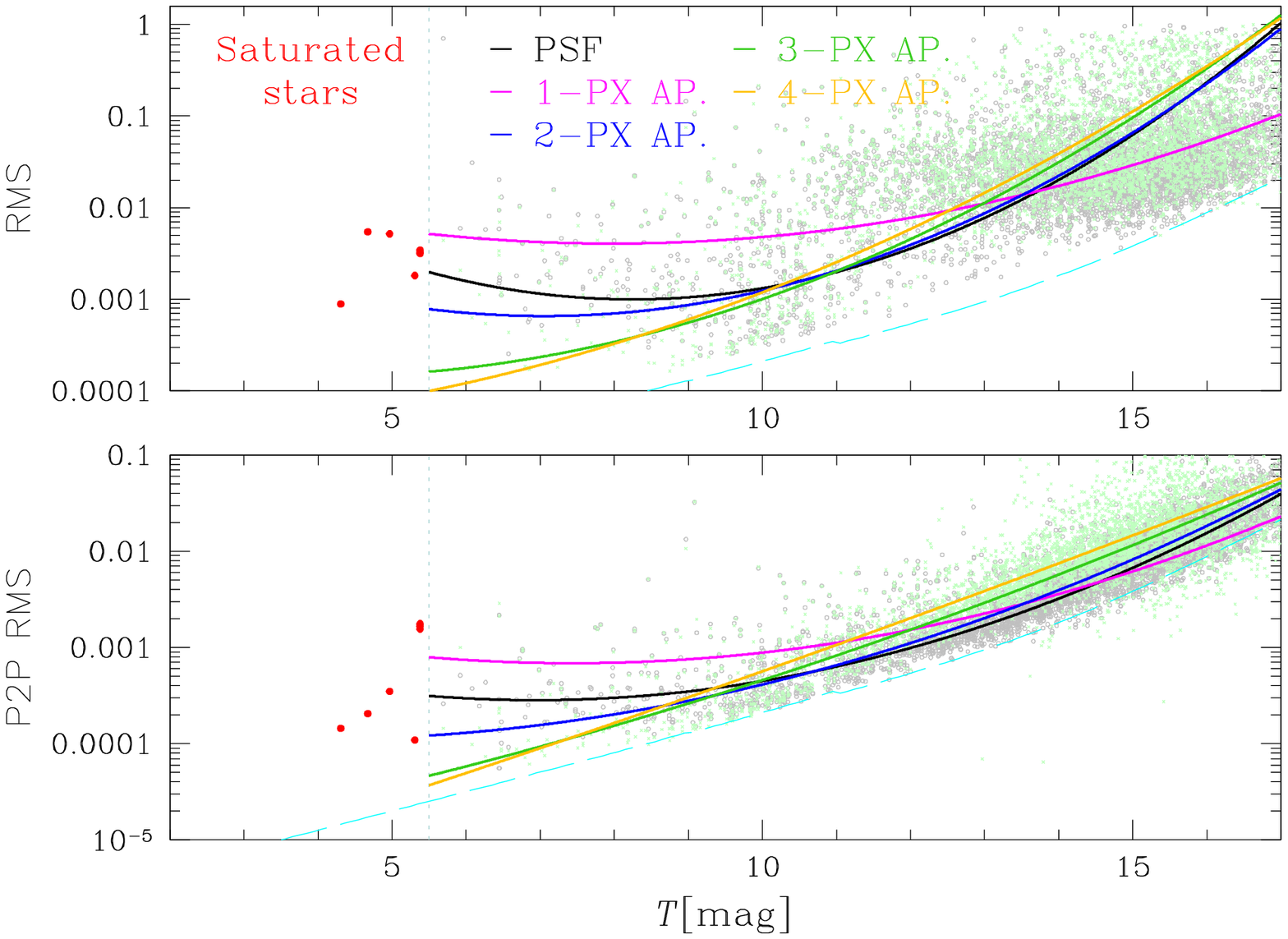}
  \caption{Photometric \texttt{RMS} and \texttt{P2P RMS} distributions
    as a function of the {\it TESS} magnitude for the light curves
    extracted in this work. Coloured lines represent the 2nd-order
    polynomial interpolation to the \texttt{RMS} distributions: black for
    PSF-fitting photometry, magenta, blue, green, and yellow for 1-px,
    2-px, 3-px, and 4-px aperture photometries, respectively. Light
    blue dashed line represents the theoretical limit. Grey dots and
    light green crosses are examples of \texttt{RMS} distributions for
    PSF-fitting and 3-px aperture photometries, respectively.  Red
    points are the saturated stars.
    \label{fig:2}}
\end{figure*}

\subsection{The input catalogue}
\label{sec:input_cat}

In this work, I analysed stars that have high probability to be
members of five associations: Chamaeleon I and II associations
(hereafter, ChI and ChII), Lupus association (Lup), Corona Australis
association (CrA), and $\gamma$ Velorum association (Vel). The
selection of likely association members was performed by using
Gaia\,DR2 (\citealt{2018A&A...616A...1G}) information, like proper
motions and parallaxes. For each association, I extracted from the
Gaia\,DR2 catalogue all the stars with $G<19$, within circular regions
(of radius $r_0$) of the sky centred in the ($\alpha_0,\delta_0$); the
values of $r_0$, $\alpha_0$, and $\delta_0$ are tabulated in
Table~\ref{tab:1}. For each region, I first analysed the proper motion
distributions of the stars with $G<15$, and I selected manually the
area of the vector-point diagram where likely association members are
located. I fitted the $\mu_{\alpha}\cos{\delta}$ and $\mu_{\delta}$
distributions with Gaussian functions and I selected all the points
within $3.5 \sigma$ from the mean values of $\mu_{\alpha}\cos{\delta}$
and $\mu_{\delta}$. I fitted the parallax ($\pi$) distribution of the
selected stars with a Gaussian, and I selected all the points within
$3.5\sigma$ from the mean value of $\pi$. I iterated the procedure 10
times, alternating the proper motion and parallax selections and using
only the stars that passed the selection criteria of the previous
iteration. An example of likely association member selection for the
CrA is illustrated in Fig.~\ref{fig:0}.  The Vel and Lup associations
are more complex systems and are formed by groups of stars having
slightly different kinematical properties; in particular, in the
vector-point diagram association stars form different close
clumps. When possible, I fitted all the single clumps with Gaussian
functions and I selected the stars as previously described. The final
catalogue given as input of \texttt{IMG2LC} contains 6629 stars.  I
cross-matched the final catalogue with the TIC v8 catalogue
(\citealt{2019AJ....158..138S}), in order to obtain photometric
information on all the stars. In particular, I included in the
catalogue (in addition to {\it TESS} and Gaia magnitudes) the
magnitudes in $B$- and $V$-Johnson bands, the 2MASS $J$, $H$, and
$K_{s}$ magnitudes (\citealt{Cutri2003}), and the infrared WISE (\citealt{Wright2010})
magnitudes $W1$ ($[3.4 \mu{\rm m}]$), $W2$ ($[4.6 \mu{\rm m}]$), $W3$
($[12 \mu{\rm m}]$), and $W4$ ($[22 \mu{\rm m}]$). Reddening values
($E(B-V)$) were extracted for each star in the input catalogue by
using the python routine
\texttt{mwdust}\footnote{\url{https://github.com/jobovy/mwdust}}
implemented by \citet{2016ApJ...818..130B}, and the
\texttt{Combined19} dustmap
(\citealt{2003A&A...409..205D,2006A&A...453..635M,2019ApJ...887...93G}).
Figure~\ref{fig:1} shows an example $G$ versus $(G-G_{\rm RP})$
colour-magnitude diagram (CMD) before (panel (a)) and after (panel
(b)) the correction for reddening for the two close associations ChI
and ChII.

\subsection{Photometric precision}

Following the same methodology as in \citetalias{2019MNRAS.490.3806N} and
\citetalias{2020MNRAS.495.4924N}, I calculated the following quality
parameters for the cotrended light curves: (i) the photometric
\texttt{RMS}, defined as the 68.27th percentile of the sorted residual
from the 3.5$\sigma$-clipped median value of the light curve; because
the simple photometric \texttt{RMS} is very sensitive to stellar
variations, I calculated the (ii) \texttt{P2P RMS}, defined as the 68.27th
percentile of the sorted residual from the median value of the vector
$\delta F_{i}= F_{i} - F_{i+1}$, where $F$ is the flux value at a given epoch
$i$.  I fitted the \texttt{RMS} and \texttt{P2P RMS} distributions
with different polynomial functions, changing the order between $n=1$
and $n=5$, to derive the best mean trend of each photometric method. I
found that, on average, the best fit was the one with
$n=2$. Figure~\ref{fig:2} shows the \texttt{RMS} (top panel) and
\texttt{P2P RMS} (bottom panel) distributions; coloured lines are the
2nd-order polynomial fits performed for each photometric method. As
done in \citetalias{2020MNRAS.495.4924N}, I used the \texttt{P2P RMS}
trends to define the best photometric method for each light curve: for
not saturated stars with $T\lesssim7.0$, I used stars extracted with
the 4-pixel aperture photometry; in the $7.0 \lesssim T \lesssim 9.5$
regime, 3-pixel aperture photometry gives the best results; for stars
having $9.5 \lesssim T \lesssim 11.0$ the 2-pixel aperture photometry
produces, on average, light curves with the lower \texttt{P2P RMS};
PSF-fitting photometry works better than the aperture photometry in
the range $11.0 \lesssim T \lesssim 14.5$; in the faint regime,
$T\gtrsim 14.5$, the best choice is the 1-pixel aperture photometry.

After this first selection, to exclude stars contaminated by different
kind of sources (bleeding columns, bad pixels, not-subtracted stars,
blended stars), I excluded all the sources for which the mean
instrumental magnitude $T_{\rm instr}$ is too different from that
expected knowing the calibrated $T_{\rm cal}$. In order to select the
best stars, I calculated the mean of the $\delta T = T_{\rm
  instr}-T_{\rm cal}$ distribution, $\bar{\delta T}$, and its standard
deviation $\sigma_{\delta T}$ and I excluded all the stars for which
$|\delta T-\bar{\delta T}| > 4\times \sigma_{\delta T}$; 4088 stars
passed all the selection criteria and have been analysed.

\begin{figure*}
  \includegraphics[bb=19 301 571 697, width=0.9\textwidth]{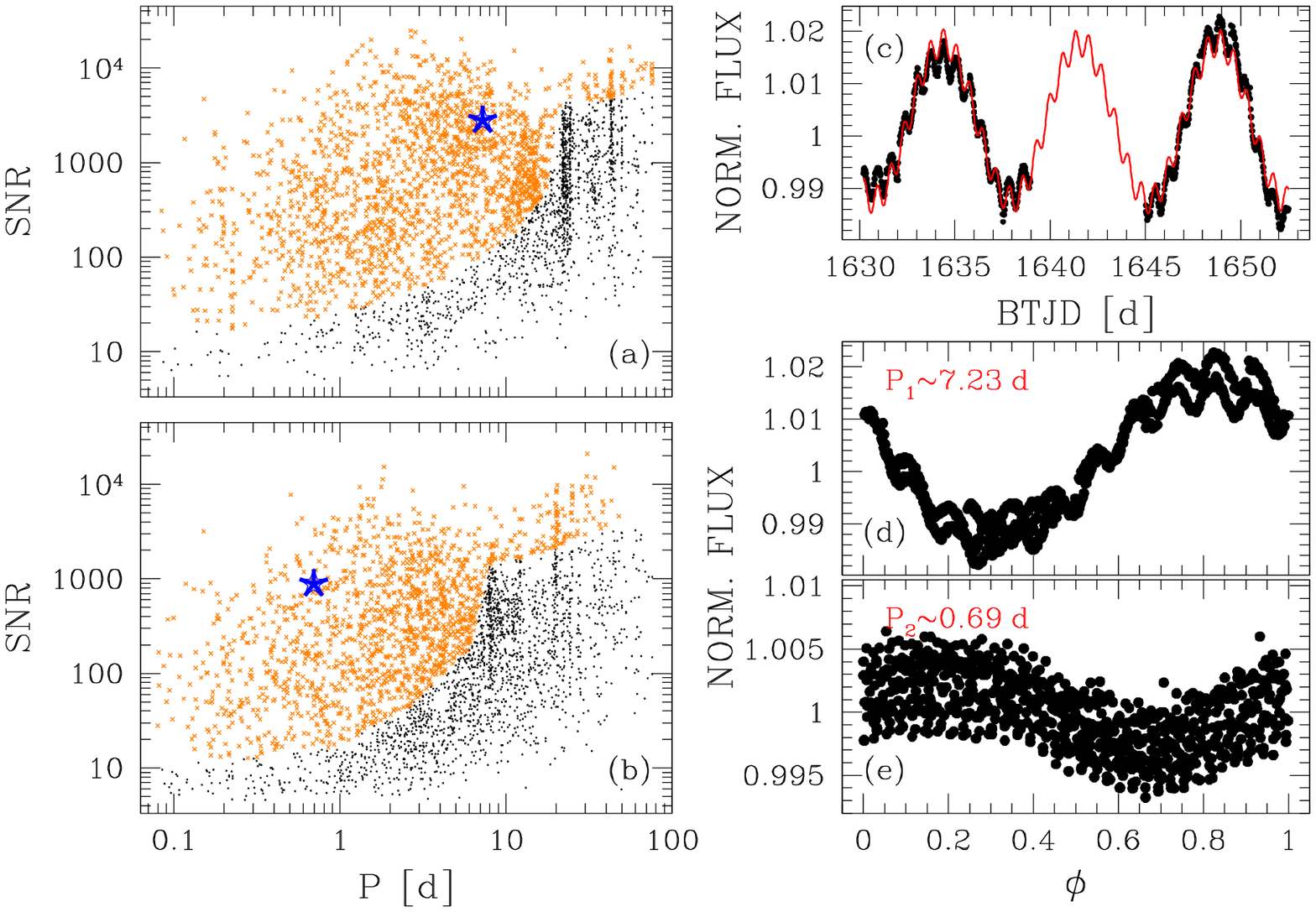} \\
  \caption{Selection of candidate variable stars. Panels (a) and (b)
    show the selection of candidate variables in the SNR versus Period
    plane, for the first and the second peak period, respectively:
    orange points are the candidate variables selected on the basis of
    their SNR; blue starred point represent the star TIC~69420071
    whose light curve is shown in panel (c). In red is the sinusoidal
    model obtained combining the two peak periods found by GLS. Panel
    (d) shows the light curve of TIC~69420071 phased adopting the
    first peak period ($\sim 7.23$~d); panel (e) shows the (whitened)
    light curve phased using the second peak period ($\sim 0.69$~d).
    \label{fig:5}}
\end{figure*}

\begin{figure*}
  \includegraphics[bb=22 449 480 711, width=0.75\textwidth]{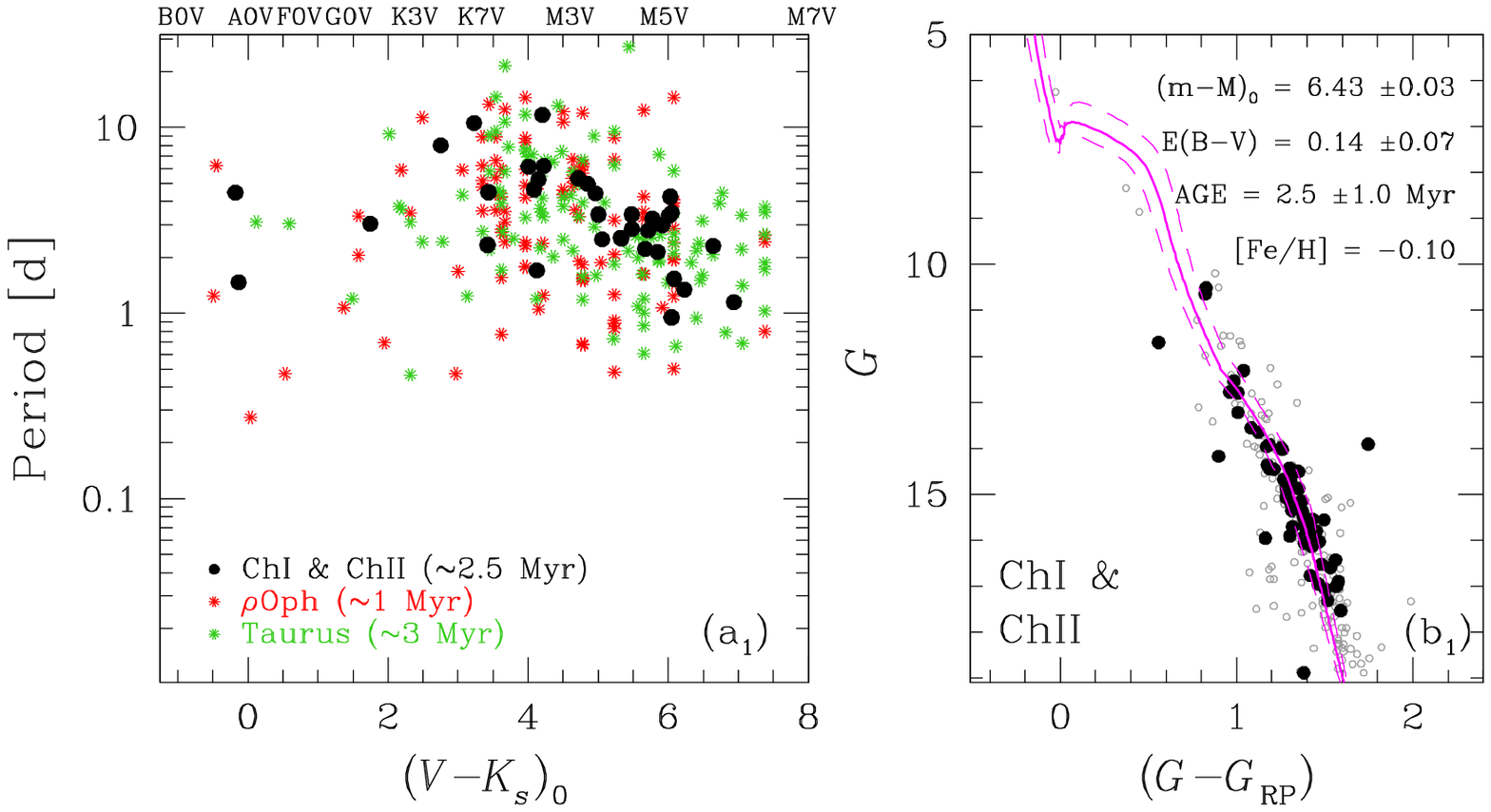} \\
  \includegraphics[bb=22 449 480 711, width=0.75\textwidth]{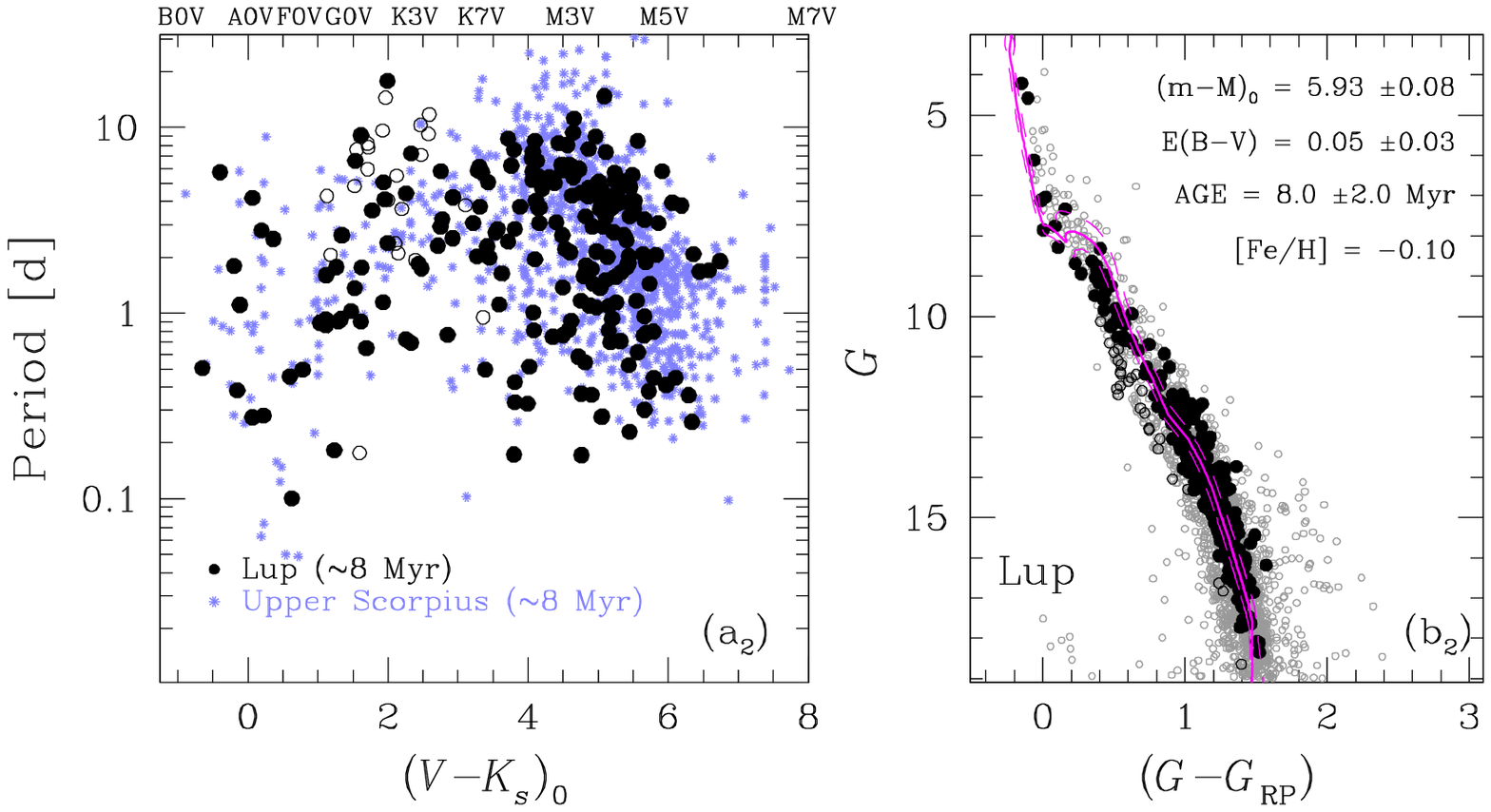} 
  \caption{Age computation of the associations studied in this work
    based on the analysis of Period-$(V-K)_0$ distributions and on the
    fit of the isochrones. Panels (a) show the Period-$(V-K)_0$ for
    one of the associations studied in this work (in black) compared
    to the distributions obtained for other associations in other
    works ($\rho$~Oph, Taurus, and Upper Sco in red, green, and
    violet, respectively,
    \citealt{2018AJ....155..196R,2020AJ....159..273R}). On the top of
    the panels (a) are reported the spectral classes as defined by
    \citet[table updated to March 2019]{2013ApJS..208....9P}. Panels
    (b) show the best fit isochrones to the $G$ versus $(G-G_{\rm
      RP})$ CMD of the single associations and the parameters used to
    obtain the fit. In this figure are shown the results for ChI, ChII
    (top panels), and Lup associations (bottom panels, empty black
    circles indicate likely field stars).
    \label{fig:4a}}
\end{figure*}

\begin{figure*}
  \includegraphics[bb=22 449 480 711, width=0.75\textwidth]{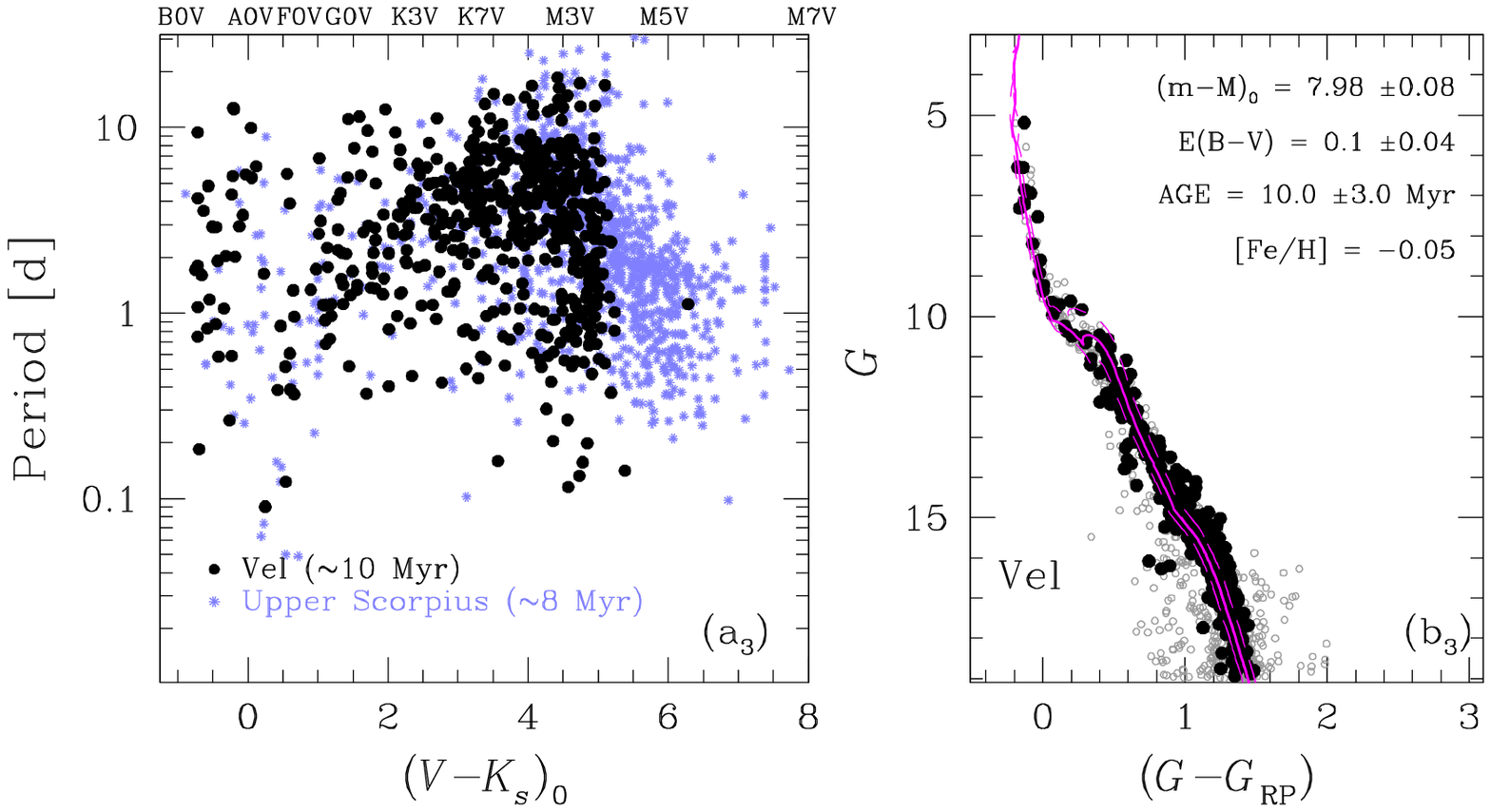}
  \includegraphics[bb=22 449 480 711, width=0.75\textwidth]{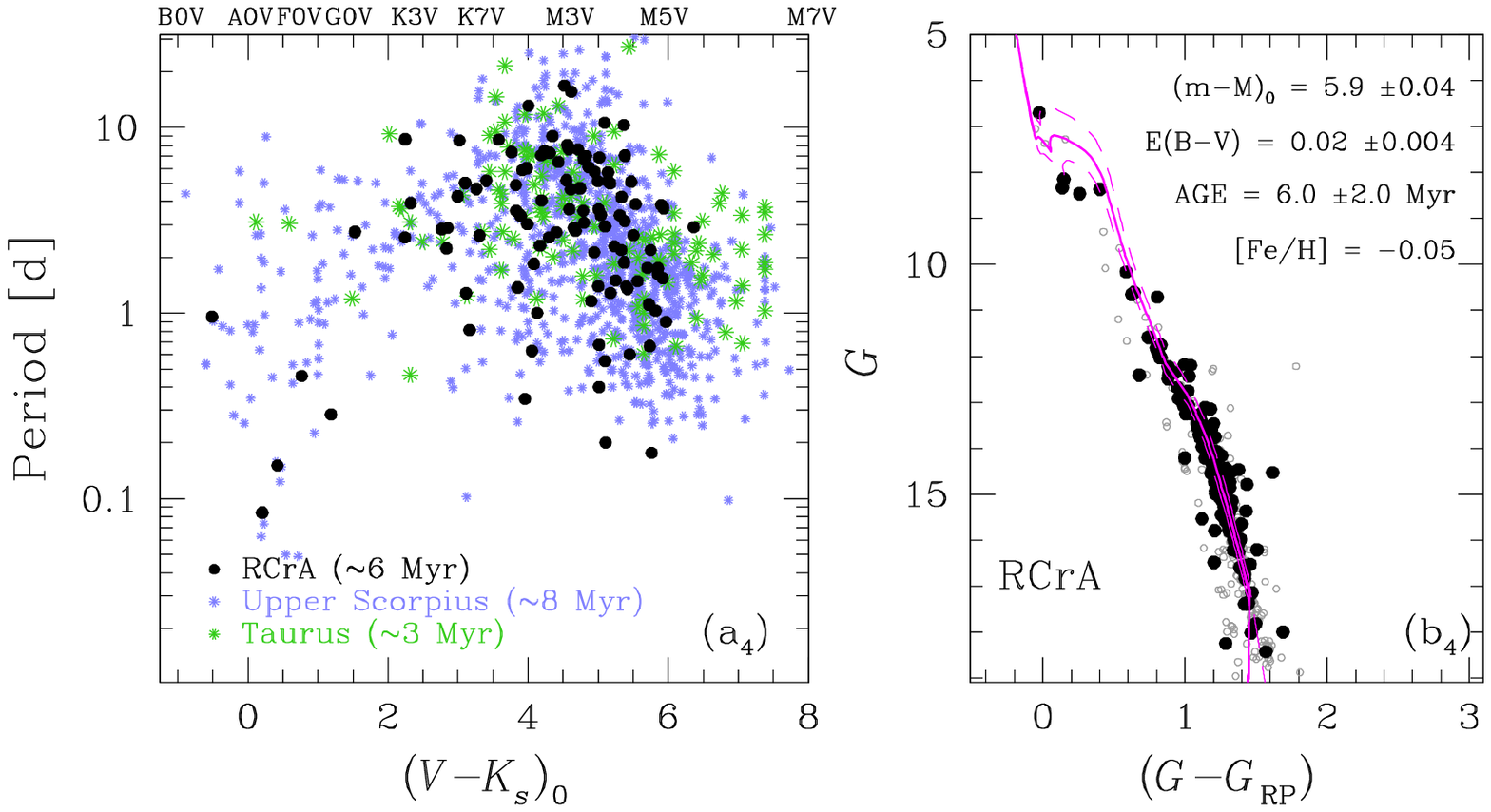}
  \caption{As in Fig.~\ref{fig:4a}, but for Vel (top panels) and CrA
    (bottom panels) associations.
    \label{fig:4b}}
\end{figure*}

\section{Stellar Variability}
\label{sec:var}

The analysis of the stellar variability of the association members is
crucial to constraint some properties of the associations.\\
In order to find periodic variable stars, I used the Generalized
Lomb-Scargle (GLS, \citealt{2009A&A...496..577Z}) routine implemented
in \texttt{VARTOOLS
  1.38}\footnote{\url{https://www.astro.princeton.edu/~jhartman/vartools.html}}(\citealt{2016A&C....17....1H})
to extract the periodograms of the light curves. After the
identification of the period associated to the most powerful peak in
the periodogram, the routine whitened the light curve and extracted
the periodogram of the light curve again to find the second strongest
peak period. The reason for this multi-period finding are: (i) some
stars present multiple signals associated to different physical
phenomena (see, e.g., \citealt{2016AJ....152..114R}), and the multi-period finding allows us to
identify the different periods; (ii) artifacts in the light curve or
effects due to the observations (sampling, temporal gaps in the light
curve, outliers, etc.) might generate a peak in the periodogram stronger
than that associated to the real physical signal coming from the
star; multiple-period finding allows us to recover the real signal.
For each light curve, I searched for periods between
0.08~d$\le P \le T_{\rm LC}$, where $T_{\rm LC}$ is the maximum
temporal baseline of the light curve. I excluded the candidate
variable stars blended with other stars in the catalogue having
similar signals using the routine \texttt{findblends} implemented in
\texttt{VARTOOLS 1.38}: this routine compares the positions of the
stars in a catalogue, their periods found by using GLS periodograms
and the amplitudes of their light curves to find blended stars.  I
used the Signal-to-Noise Ratio (SNR) parameter to isolate the
candidate variable stars following the method described in
\citet{2015MNRAS.447.3536N} and shown in Fig.~\ref{fig:5}: I divided
the SNR distribution in intervals of $\delta P= 1$~d, and I computed
the 3.5$\sigma$-clipped mean and standard deviation of the SNR values
inside each bin. I interpolated the points $3 \sigma$ above the mean
SNR values with a spline, and I considered as candidate variables the
points above the interpolated line (orange points in panels (a) and
(b) of Fig.~\ref{fig:5}). I applied this procedure both to the SNR
distributions associated to the first peak of the periodograms and to
the SNR associated to the second peak periods; I considered as
candidate variables the stars selected in both the sample (2230
stars). Finally, I visually inspected the phased light curves to
assign to each candidate variable star the corrected period (or both
the periods if the star have multiple periods), or to discard it
because false positive. Panels (c), (d), and (e) show an example of
light curve of a star characterised by multiple periods.  The
final list of periodic variable stars contains 1260 stars, 28 of them
have multiple periods.  A list of periodic variable stars used in this
work is available electronically. The description of the columns are
reported in Table~\ref{tab:a1}.

\subsection{Period-colour distribution analysis}
\label{sec:PC}

Because of their young age and of the low number of members, the
estimation of the association ages based on the use of theoretical
models is not immediate.  By using gyrochronology, i.e. the method for
the estimation of the age based on the analysis of stellar rotation
and magnetic braking
(\citealt{2003ApJ...586..464B,2007ApJ...669.1167B}), it is possible to
constraint the age of the associations studied in this work.


In this work, I combined the period-colour distribution analysis
and the CMD isochrone fitting, in order to constraint the age of each
association and use this parameter in the characterisation of the
candidate transiting exoplanets (Sect.~\ref{sec:cand_exo}). In a first
step, I found a raw estimation of the ages of the five associations
comparing the period-$(V-K_{s})_0$ dereddened colour distributions of
the associations studied in this work with that obtained by
\citet{2018AJ....155..196R} for the $\rho$~Oph ($\sim 1$~Myr) and
Upper Sco ($\sim 8$~Myr) associations, and with the period-colour
analysis performed by \citet{2020AJ....159..273R} for the Taurus
association ($\sim 3$~Myr). I used this first guess on the age to
perform a fit of the isochrones and extract the age of each
association. The associations studied in this work have a slightly
subsolar metallicity ([Fe/H]$\sim -0.10$~-~$-0.05$, see, e.g.,
\citealt{2006A&A...446..971J,2014A&A...567A..55S,2014A&A...568A...2S};
Table~\ref{tab:1}). For the isochrones fitting I used two sets of
metallicities: for ChI, ChII and Lup associations I used isochrones with
[Fe/H]=$-0.10$, while for Vel and CrA isochrones with [Fe/H]=$-0.05$.

\noindent
(i) \textbf{ChI \& ChII associations.} Because of the low number of
members and because the two associations are almost coeval and at the
same distance ($\sim 195$~pc), I analysed their period-colour
distributions together. Panel (a$_1$) of Fig.~\ref{fig:4a} shows the
$P$ vs $(V-K_{s})_0$ distributions for ChI \& ChII (black points)
compared to that of $\rho$~Oph (red points) and Taurus (green points)
associations: in the four associations low-mass slow-rotator ($P \sim
2$--$10$\,d) stars prevail. It means that all the associations are
almost coeval. The associations $\rho$~Oph and Taurus are very young
with ages between $\sim 1$ and $\sim 3$~Myr
(\citealt{2018AJ....155..196R,2020AJ....159..273R}). I used this
information to constraint the fit of the isochrones shown in panel
(b$_1$). I computed the median reddening $E(B-V)$ and the median
distance modulus\footnote{By using the Gaia\,DR2 parallaxes corrected
  for the $-30~\mu$mas offset found by \citet{2018A&A...616A...2L}}
$(m-M)_0$ of the stars that belong to ChI and ChII associations, and I
performed a $\chi^2$-fit of a set of PARSEC (PAdova and TRieste Stellar
Evolution Code,
\citealt{2002A&A...391..195G,2012MNRAS.427..127B,2017ApJ...835...77M})
isochrones
\footnote{\url{http://stev.oapd.inaf.it/cgi-bin/cmd}} with ages
that run from 1 to 5 Myr, in step of 0.5~Myr, to the $G$ versus $(G-G_{\rm RP})$
CMD, as done by \citet[I refer the reader to this work for a detailed
  description of the fit procedure]{2015MNRAS.451..312N}. I found that
the best fit is associated to an age of $2.5 \pm 1.0$~Myr.

\noindent
(ii) \textbf{Lup association.} Even if I performed strict selections
on proper motions and parallaxes for the groups that form the complex
Lup association, some field stars are still present in the
catalogue. In the analysis of variable stars, I excluded these likely
field stars on the basis of their colours and magnitudes. Panel
(a$_2$) of Fig.~\ref{fig:4a} shows the period-colour distribution of
variable stars in the Lup association compared to that derived by
\citet{2018AJ....155..196R} for Upper Sco stars (empty black circles
are the likely field stars). The two distributions are very similar,
with a scattered sequence of AFGK stars which become slower as the
mass decreases, and a well populated sequence of M stars, whose
periods decrease from early to late spectral types. The age of Upper
Sco is $\sim 8$-$10$~Myr
(\citealt{2012ApJ...746..154P,2018AJ....155..196R}): starting from
this constraint, I performed a fit of isochrones (with ages between 5
and 15~Myr) to the Lup association CMD (panel (b$_2$) of
Fig.~\ref{fig:4a}). The best fit is obtained for an age of $8.0 \pm
2.0$~Myr.

\noindent
(iii) \textbf{Vel association.} Because the Vel association is further
away than the other associations studied in this work, low mass stars
have luminosities over the {\it TESS} magnitude limit and the result
is that the period-colour distribution is cut on the red part, as
shown in panel (a$_3$) of Fig.~\ref{fig:4b}. Even if the sequence of M
dwarfs is incomplete, the sequence formed by AFGK stars (which periods
increases with the colour) and part of the sequence of M-type stars
are very similar to that of the Upper Sco association. As done for the
Lup association, I used a set of isochrones with ages between 5 and
15~Myr to search the age that gives the best-fit. I found an age of
$10.0 \pm 3.0$~Myr, as shown in panel (b$_3$) of Fig.~\ref{fig:4b}

\noindent
(iv) \textbf{CrA association.} The period-colour distribution of
variable stars in the CrA association is shown in panel (a$_4$) of
Fig.~\ref{fig:4b}, compared to the distributions of Taurus and Upper
Sco stars. The distribution of the periods of the M stars in the CrA
association follows that of the M stars in the Upper Sco, with late
type stars that are faster rotators than early M dwarfs. Unfortunately
few stars with spectral types earlier than M populate the
period-colour distribution, and a direct comparison of this part of
the distribution is not possible. Using these information, I performed
a fit of the CMD using isochrones with ages between 1 and 15~Myr; I
found the best fit for an age of $6.0 \pm 2.0$~Myr (panel (b$_4$) of
Fig.~\ref{fig:4b}).

\begin{figure*}
  \includegraphics[bb=17 145 590 717, width=0.9\textwidth]{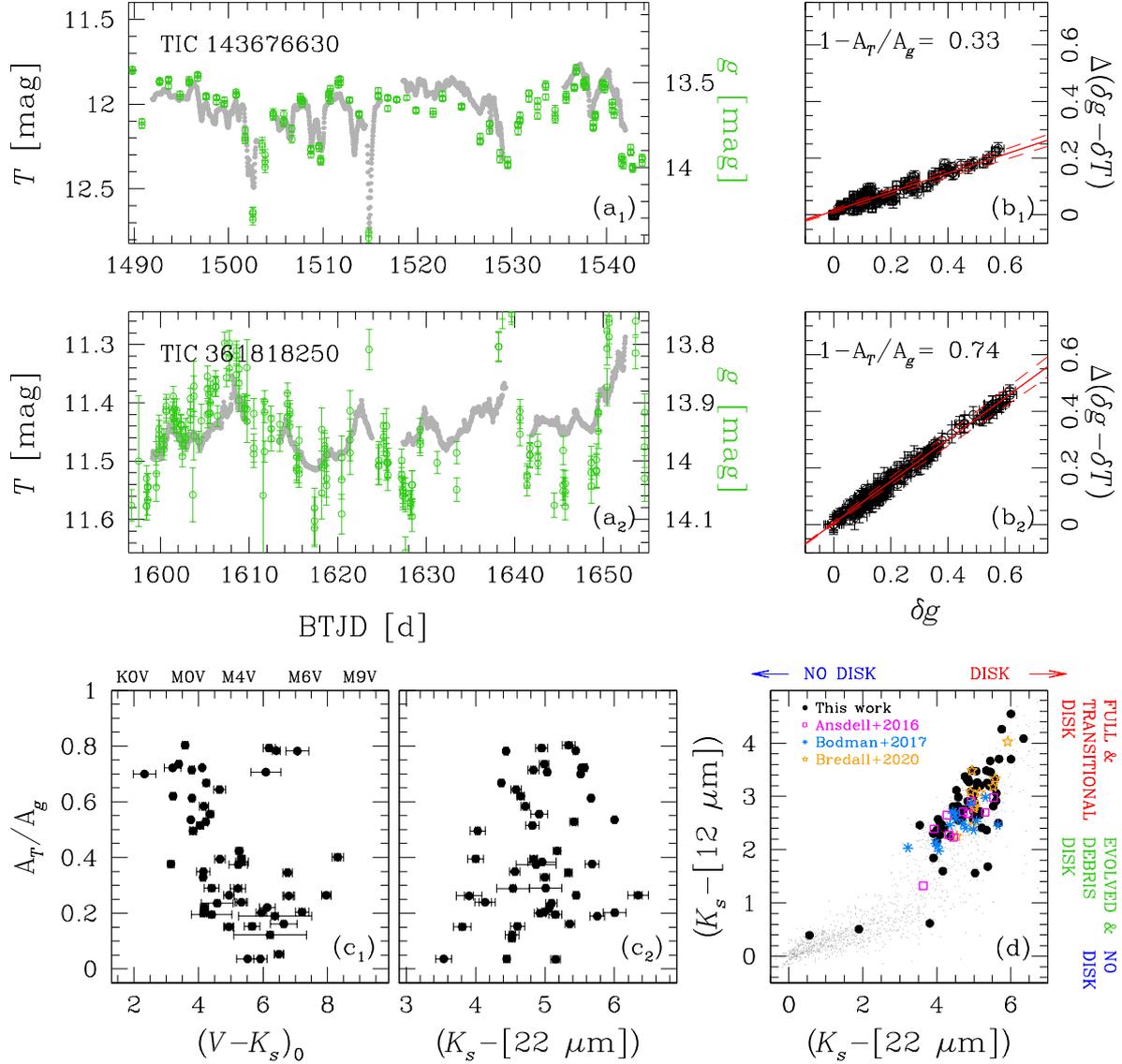}
  \caption{Analysis of the dippers. Panels (a) show two example of
    light curves of dipper stars obtained using {\it TESS} data (grey
    points) and ASAS-SN $g$-band data (green points); the magnitude
    limits on the y-axis cover the same interval for {\it T} and {\it
      g} magnitudes, in order to directly compare light curves
    obtained with different instruments. Panels (b) show the $\Delta(
    \delta g - \delta T)$ versus $\delta g$ diagram: red lines are the
    best least squares fit. The slope of the fitted line is correlated
    to the reddening caused by the circumstellar disk and is related
    to the size of the grains that form the dust. Panel (c$_1$) is the
    $A_{T}/A_{g}$ versus the dereddened colour index $(V-K_{s})_0$
    diagram: on the top are reported the spectral classes as defined
    by \citet[table updated to March 2019]{2013ApJS..208....9P}. Panel
    (c$_2$) shows $A_{T}/A_{g}$ versus the infrared excess indicator
    $(K_{s}-[22\mu{\rm m}])$. Panel (d) is the $(K_{s}-[12\mu{\rm
        m}])$ versus $(K_{s}-[22\mu{\rm m}])$ colour-colour diagram as
    reported by \citet{2012ApJ...758...31L}: the different
    evolutionary stages of the disk are reported on the right. Grey
    points are all the stars in the input catalogue adopted in this
    work, black filled circles are the dipper stars identified in this
    work, magenta empty squares, azure asterisks, and orange stars are
    the dippers identified by \citet{2016ApJ...816...69A},
    \citet{2017MNRAS.470..202B}, and \citet{2020MNRAS.tmp.1773B},
    respectively.
    \label{fig:3}}
\end{figure*}

\subsection{Dipper stars and disk properties}

Young stellar systems, like the associations analysed in this work,
host low mass ($\lesssim 1 M_{\sun}$) T Tauri-like ``dipper'' stars
surrounded by circumstellar disks.  Dipper stars are young stellar
objects (YSOs) that show dimming events (periodic or not) in their
light curves, probably caused by the dust located in the inner regions
of a circumstellar disk that ``transits'' on the stellar disk
(\citealt{2017MNRAS.470..202B}). The luminosity of these stars usually
decreases between few percent to $>1$\,magnitude, on timescales
between few-hours and about 1 day. \\
Characterise dipper stars and their disks in young associations with
different ages is essential to understand how they evolve and which
are the cleaning timescales of disks, allowing us to constraint the
models on the planet formation.

To date, few ground-based surveys have been performed to study these
objects (see, e.g., \citealt{2010ApJS..191..389C,2011ApJ...733...50M});
in the last years data from telescopes in space  ({\it K2/Kepler} and {\it
  CoRot}), gave a great contribution to the analysis of dipper stars
(see,
e.g., \citealt{2015AJ....149..130S,2016ApJ...816...69A,2017ApJ...848...97R,2018AJ....156...71C}),
but these missions had very limited sky coverage. Recently,
\citet{2020MNRAS.tmp.1773B} characterised 11 stars in the Lupus
region, combining ground-based (ASAS-SN) and space-based ({\it
  TESS}) data. In fact, {\it TESS} is offering a unique opportunity to
study with an high photometric accuracy the evolution of the light
coming from these stars, over a long time baseline ($\gtrsim
1$\,month).

In this section I describe the procedure I followed to search and
characterise the dippers among the association members for which I
extracted the {\it TESS} light curves.

In order to search for dipper stars, I used three different metrics:
(i) the \texttt{RMS}, sensitive to the scatter of
the light curve; (ii) the peak-to-peak variability metric ($\nu$), as
defined by \citet{2017MNRAS.464..274S}, that is sensitive to the
variability of the star in general; (iii) the Flux Asymmetry ($M$),
defined by \citet{2014AJ....147...82C} and
\citet{2018AJ....156...71C}, sensitive to fading/brightening events in
the light curve. First, I divided the \texttt{RMS} distribution in bin
of 1.0\,$T$-magnitude, and, within each interval, I computed the mean
${\bar{\tt RMS}}$ and the standard deviation of the $\sigma_{\bar{\tt
    RMS}}$; I interpolated the ${ \bar{\tt RMS}+ 3\times
  \sigma_{\bar{\tt RMS}}}$ points with a cubic spline and I selected
all the sources above the interpolation. I performed the same
procedure using as parameter $\nu$, and I discarded all the points
that were not selected in \texttt{RMS} and $\nu$ selections and having
$M<-0.25$. I visually checked the light curves of the 652 stars that
passed the selection, identifying 71 candidate dippers ($>90$~\%
associated to stars of spectral type K and M).

Following the procedure adopted by \citet{2020MNRAS.tmp.1773B}, I used
the All-Sky Automated Survey for SuperNovae (ASAS-SN,
\citealt{2014ApJ...788...48S,2017PASP..129j4502K}) $g$-band light
curves to calculate the ratio between the extinction coefficient
$A_{T}$ in $T$-band and that in $g$-sloan band, $A_{g}$. This quantity
is strictly linked to the grain size of the dust that surrounds the
star. Defining $\delta T$ and $\delta g$ the dimming of the {\it TESS}
and ASAS-SN light curve, the quantity $\Delta(\delta g-\delta T)$
represents the reddening $E(g-T)=A_{g}-A_{T}$ caused by the dust. Therefore:
\begin{equation}
  \frac{\Delta(\delta g-\delta T)}{\delta g} = \frac{A_{g}-A_{T}}{A_{g}} = 1-\frac{A_{T}}{A_{g}}
\end{equation}
and the quantity $A_{T}/A_{g}$ can be inferred measuring the slope of
the $\Delta(\delta g-\delta T)$-$\delta g$ relation. I downloaded from
the ASAS-SN archive\footnote{\url{https://asas-sn.osu.edu/}} the
$g$-band light curves for all the dippers having $T<14.5$ (53 stars),
with a baseline that covers the {\it TESS} observational period of the
first year of mission. Panels (a) of Fig.~\ref{fig:3} show two examples
of light curves of dippers observed for two consecutive {\it TESS}
sectors by {\it TESS} (grey points) and ASAS-SN (green points).  I
extracted the relationship between $\Delta(\delta g-\delta T)$ and
$\delta g$ splitting the light curves in sub-sectors, each one ending
with the {\it TESS} down-link of the data (about every $13.5$ days): in
this way I avoid (2nd-order) systematic effects due to the variation
of the photometric zero-point between the first and second part of a
sector. Panels (b) shows $\Delta(\delta g-\delta T)$ as a function of
$\delta g$ for the two stars showed in panels (a): I performed a
linear least-squares fit to the data of each sub-sector to obtain the
slope $(1-A_{T}/A_{g})_i$, with $i=1,...N_{\rm ssec}$ is $i$-th
sub-sector, and, finally, I averaged all the slopes. The fits obtained
with the mean slope are shown in panels (b) of Fig.~\ref{fig:3} (red
lines). The catalogue of the identified dippers and of the
$A_{T}/A_{g}$ values is released as electronic material;
Table~\ref{tab:a2} reports the description of this catalogue.

The ratio $A_{T}/A_{g}$ gives information about the size of the grains
that form the surrounding disk: if the dust is dominated by small
grains, the quantity $A_{T}/A_{g}$ will be smaller than the case in
which the grains have large size; if the size of the grains are larger
than the wavelengths in which the {\it TESS} observations were
performed ($\lambda_{\rm central}\sim 800$\,nm), the ratio
$A_{T}/A_{g} \to 1$, and the reddening $E(g-T)=A_{g}-A_{T}
\to 0$.

Panel (c$_1$) of Fig.~\ref{fig:3} shows the $A_{T}/A_{g}$ as a
function of the de-reddened colour $(V-K_{s})_0$: the two quantities
are slightly correlated (Pearson coefficient $\sim -0.4$), with, on
average, earlier type stars having disks formed by larger grains.
\citet{2020MNRAS.tmp.1773B} found a weak relation between the grain
sizes and the infrared excess measured with the colour
$(K_{s}-[22\mu{\rm m}])$, with the infrared excess that inversely
decreases with the dimension of the grains. Panel (c$_2$) of
Fig.~\ref{fig:3} illustrates the distribution of the $A_{T}/A_{g}$
measured in this work as a function of the infrared excess
$(K_{s}-[22\mu{\rm m}])$: it shows that there is not a clear
correlation between the two quantities (Pearson coefficient: $\sim
0.3$), and it does not confirm what found by
\citet{2020MNRAS.tmp.1773B}.

The presence of a disk around the dippers found in this work is also
confirmed by the analysis of the excess emission in infrared shown in
panel (d) of Fig.~\ref{fig:3}. In fact, the evolutionary stage of a
disk can be inferred comparing the stellar luminosity in the 2MASS
$K_{s}$ band with its WISE infrared magnitudes. As reported by
\citet{2012ApJ...758...31L}, in a $(K_{s}-[12\mu{\rm m}])$ versus
$(K_{s}-[22\mu{\rm m}])$ colour-colour diagram, stars with
$(K_{s}-[22\mu{\rm m}])\gtrsim 3.2$ are surrounded by {\it full}/{\it
  transitional} disks\footnote{I refer the reader to
  \citet{2012ApJ...758...31L} for a detailed description of the
  different evolutionary stages of the disks}, while {\it evolved} and
      {\it debris} disks are located in the area defined by
      $(K_{s}-[22\mu{\rm m}])\gtrsim 3.2$ and $(K_{s}-[12\mu{\rm
          m}])\gtrsim 0.5$. Stars with small values of the
      colour-colour indexes ($\sim 0$) have no disk. Panel (d) of
      Fig.~\ref{fig:3} confirms that the large part of YSOs-dippers
      found in this work have {\it full} or {\it transitional} disks;
      about 5 objects have {\it evolved} or {\it debris} disks. For
      completeness, I also report  the results by
      \citet{2016ApJ...816...69A}, \citet{2017MNRAS.470..202B}, and
      \citet{2020MNRAS.tmp.1773B}.  I found that about half of dippers
      found in this work are located in the ChI and ChII associations
      (23 and 14, respectively) and the other half in the Lup, CrA,
      and Vel associations (19, 5, and 10, respectively). Considering
      the number of stars for which I studied the light curves, I
      found that in the very young associations ChI and ChII ($\sim
      2.5$~Myr) there is a high fraction of dippers ($\sim 12$\% and
      $\sim 28$\%, respectively), while the the fraction of dippers
      decreases considering the other older associations: Lup and CrA
      ($\sim 6$-$8$~Myr) associations contain $\sim 1-2$~\% of dipper
      stars, while in the Vel association ($\sim 10$~Myr) only $\sim$0.4~\%
      of analysed stars are dippers, confirming that disks around
      low-mass stars ($\sim$0.1--0.5 $M_{\sun}$) survive up to $10$~Myr
      (see, e.g., \citealt{2009ApJ...705.1646C,2012ApJ...758...31L}).

\begin{figure*}
  \includegraphics[bb=19 166 570 695, width=0.9\textwidth]{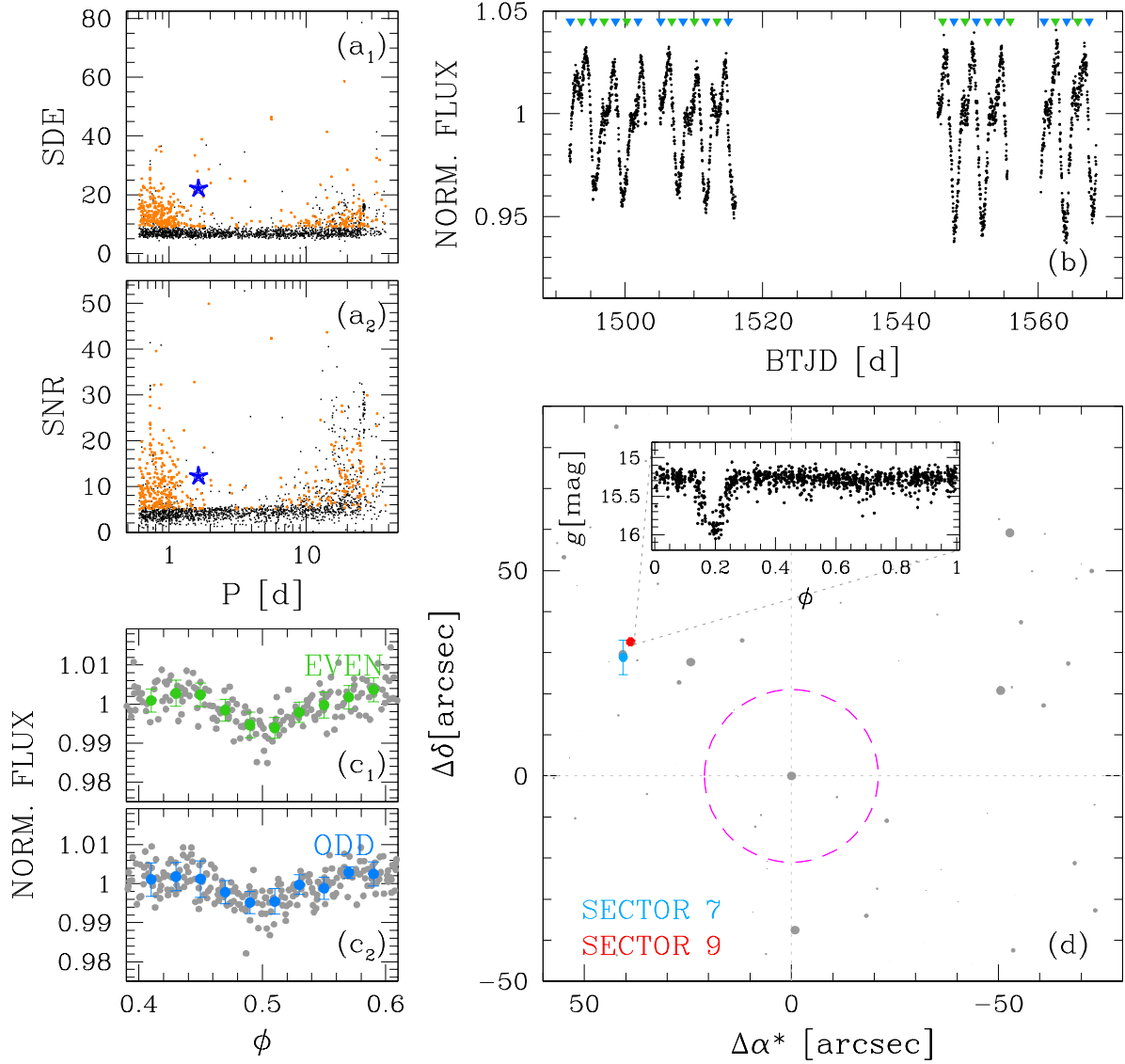}
  \caption{Overview on the selection of candidate transiting
    objects. Panels (a) show the SDE and SNR versus Period
    distributions obtained with the TLS routine; blue star represents
    TIC~143777072. Panel (b) shows the light curve of TIC~143777072
    observed in Sectors 7 and 9; green and azure triangles indicate
    the even and odd transits, respectively. Panels (c) are the phased
    even and odd transits: green/azure points are the median flux
    values calculated in bins of width 0.01. Panel (d) is the analysis
    of in/out-of-transit difference centroid: in (0,0) is located
    TIC~143777072; magenta circle has the same size of the photometric
    aperture adopted in the light curve analysis; light blue and red
    points are the mean centroids calculated for sectors 7 and 9,
    respectively, and indicate that the transit events happen on the
    suspected eclipsing binary TIC~143777056, whose ASAS-SN $g$-band
    light curve is plotted in the inset panel.  \label{fig:6}}
\end{figure*}

\section{Candidate exoplanets}
\label{sec:cand_exo}
I searched for candidate exoplanets among the association members by
using the procedure described in
\citetalias{2020MNRAS.495.4924N}. In order to search for
transits in the light curves, stellar variability must be removed from
them. I modelled the variability of each light curve interpolating to
it a 5th-order spline defined on $N_{\rm knots}$. I considered three
different grids of knots (with knots every 4.0~h, 6.5~h, and 13.0~h)
to better model the light curve of short- and long-period variable
stars and also to avoid the flattening of transits whose duration is
longer than 4 or 6.5 hours. I removed bad photometric measurements
from the flattened light curves by clipping away all the outliers $4
\sigma$ above the median flux, and discarding all the points with
\texttt{DQUALITY>0} and values of the local background $5\sigma$ above
the mean background value.

Adopting the routine developed by \citet{2019A&A...623A..39H}, I
extracted the Transit-fitting Least squares (TLS)
periodograms\footnote{\texttt{TLS v.1.0.24};
  \url{https://github.com/hippke/tls}} of the flattened and
``cleaned'' light curves. I searched for transits with period
$0.6$\,d$\le P<0.5 \times T_{\rm LC}$, where $T_{\rm LC}$ is the
maximum temporal interval covered by the light curve. I performed a
first selection of candidate transiting objects on the basis of four
parameters extracted by the TLS routine: (i) the signal detection
efficiency (SDE); (ii) signal-to-noise ratio (SNR); (iii) the
significance between odd and even transits ($\sigma_{\rm odd-even}$);
(iv) the mean depth of the transits ($\delta_{\rm t}$). I selected as
candidates all the stars having: (i) SDE$\ge 9$ (panel (a$_1$) of
Fig.~\ref{fig:6}); (ii) SNR$\ge 5$ (panel (a$_2$) of
Fig.~\ref{fig:6}); (iii) $\sigma_{\rm odd-even}$<3; (iv) $\delta_{\rm
  t}$<10\,\%. I visually inspected the light curves that passed the
selections to check the odd/even transit depths (panels (c) of
Fig.~\ref{fig:6}), the presence of secondary eclipses, and to exclude
false positives due to the presence of artefacts in the light curves.
For each sector, I applied this procedure to the light curves flattened
by using the three different grids of knots previously described. I
repeated this procedure considering as a first step each sector
independent from the others, and then considering the stacked light
curves of the stars observed in more than one sector: in this way I
avoided that possible artefacts in (or different photometric precision
of) the light curves of stars observed in more than one sector
decrease the detection efficiency of the TLS routine. The number of
candidates that passed this first selection is 48.

I performed a series of vetting tests on the light curves of these 48
candidate transiting objects: (i) inspection of the light curves
obtained with different photometric apertures in order to check
changes in the transit depths due to a close eclipsing binary; (ii)
check of the light curves phased with a period of 0.5$\times$,
1$\times$, and 2$\times$ the period found by the TLS routine, in order to search for secondary eclipses; (iii)
comparison between the binned even/odd folded transits, in order to
check if the depth of the transits are in agreement within the errors;
(iv) analysis of the in/out-of-transit difference centroid to check if
the transit events are associated to a close contaminant. I refer the
reader to \citetalias{2019MNRAS.490.3806N} and
\citetalias{2020MNRAS.495.4924N} for a detailed description of the
vetting tests. Figure~\ref{fig:6} shows an overview of the main steps of
the vetting procedure: the not-flattened light curve and the position
of odd/even transits of the candidate TIC~143777072 are shown in panel
(b); panels (c) illustrate the comparison between odd and even
transits: because the mean depths of the transits agree within the
errors, the candidate passed this test; panel (d) shows the analysis
of the in/out-of-transit difference centroid: in both the sectors in
which the star was observed, the mean centroid is not located on the
candidate but on a star (TIC~143777056) located at $\sim 50$\,arcsec
from the target. I checked the light curve of the contaminant in the
ASAS-SN archive. The result is reported in the inset of the panel (d):
the contaminant is confirmed to be an eclipsing binary (depths of the
primary eclipse $\delta_{\rm T}\sim 0.7$~mag in $g$-band).

After the vetting procedure, 9 objects of interest (PATHOS-35--43)
belonging to two associations (Lup and Vel) survived. Among
them there are two {\it TESS} Objects of Interest
(TOI)\footnote{\url{https://tess.mit.edu/toi-releases/go-to-alerts/}}
released by the {\it TESS} team (TOI-508=PATHOS-36, TOI-831=PATHOS-41).

\subsection{Stellar parameters}

In order to extract physical parameters of the transiting objects from
the light curves of their host, some star parameters, like
temperature, mass, and radius, are mandatory. I extracted the
information for each star that hosts a candidate transiting exoplanet
by fitting isochrones to the CMDs of the associations. Stellar
parameters are derived interpolating the colour and the magnitude of
the host star on the isochrone. For the isochrone fitting, I used the
set of PARSEC isochrones, the distance modulus, the
reddening, and the metallicity adopted in Sect.~\ref{sec:PC}, and the
ages derived in the same section.  Stellar parameters for the host of transiting objects
 are reported in Table~\ref{tab:2}. These
information are used for the transit modelling, as described in the
next section.

\begin{table*}
  \caption{Star parameters and priors for the modelling.}
  \resizebox{1.01\textwidth}{!}{
  \begin{tabular}{l c l S[table-format=4.0(4)] S[table-format=4.0(4)] S[table-format=3.0(1)] c c c l c c c c}
  \hline
  TIC & PATHOS & Assoc. & {$\alpha$} & {$\delta$} & {$T$} & $R_\star$ & $M_\star$ & $\rho_\star$ & \multicolumn{1}{c}{Period} & $T_0$ & LD$_{c1}$ & LD$_{c2}$ & $df$  \\
  &        &            & {(deg.)}   &  {(deg.)}  &  {(mag.) }   & ($R_{\sun}$) & ($M_{\sun}$) & ($\rho_{\sun}$) & \multicolumn{1}{c}{(d)}    & (BTJD) &         &       &    \\
  \hline
0081353413 & 35 & Vel & 119.8064 & -49.9737 &   6.5 &  $1.04 \pm 0.15$ &  $0.72 \pm 0.05$ & $0.65\pm0.17$   & $\mathcal{U}(1.9,   2.1)$ &   $\mathcal{U}(1492.0, 1493.0)$ &  $0.38\pm0.10$   & $0.23\pm0.10$ &    $0.43\pm0.05$   \\
0081419525 & 36 & Vel & 123.0857 & -47.0900 &  14.1 &  $1.67 \pm 0.01$ &  $2.35 \pm 0.05$ & $0.50\pm0.01$   & $\mathcal{U}(1.4,   1.6)$ &   $\mathcal{U}(1517.0, 1519.0)$ &  $0.15\pm0.10$   & $0.19\pm0.10$ &    $0.16\pm0.01$   \\
0095003423 & 37 & Lup & 123.1082 & -46.1091 &   9.5 &  $1.07 \pm 0.10$ &  $0.77 \pm 0.04$ & $0.62\pm0.10$   & $\mathcal{U}(1.4,   1.6)$ &   $\mathcal{U}(1630.0, 1631.0)$ &  $0.44\pm0.10$   & $0.33\pm0.10$ &    $0.05\pm0.01$   \\
0123755508 & 38 & Vel & 115.8048 & -47.7691 &  12.5 &  $1.26 \pm 0.10$ &  $1.13 \pm 0.03$ & $0.56\pm0.08$   & $\mathcal{U}(15.0, 16.0)$ &   $\mathcal{U}(1492.0, 1493.0)$ &  $0.41\pm0.10$   & $0.36\pm0.10$ &    $0.11\pm0.02$   \\
0238235254 & 39 & Vel & 229.0334 & -38.8754 &   8.8 &  $3.90 \pm 0.10$ &  $8.25 \pm 0.17$ & $0.14\pm0.01$   & $\mathcal{U}(5.4,   5.7)$ &   $\mathcal{U}(1497.0, 1498.0)$ &  $0.02\pm0.20$   & $0.02\pm0.20$ &    $0.50\pm0.05$   \\
0238379370 & 40 & Vel & 242.5224 & -31.4072 &  11.4 &  $1.36 \pm 0.11$ &  $1.23 \pm 0.05$ & $0.49\pm0.07$   & $\mathcal{U}(27.0, 27.3)$ &   $\mathcal{U}(1511.5, 1512.0)$ &  $0.38\pm0.10$   & $0.34\pm0.10$ &    $0.04\pm0.01$   \\
0307610438 & 41 & Lup & 242.1278 & -38.4742 &  10.0 &  $1.94 \pm 0.12$ &  $1.55 \pm 0.12$ & $0.21\pm0.03$   & $\mathcal{U}(1.4,   1.6)$ &   $\mathcal{U}(1601.5, 1602.5)$ &  $0.33\pm0.10$   & $0.31\pm0.10$ &    $0.01\pm0.01$   \\
0374732772 & 42 & Lup & 120.7446 & -49.0183 &  12.2 &  $1.43 \pm 0.09$ &  $1.26 \pm 0.04$ & $0.44\pm0.05$   & $\mathcal{U}(17.5, 19.5)$ &   $\mathcal{U}(1631.0, 1632.0)$ &  $0.39\pm0.10$   & $0.35\pm0.10$ &    $0.05\pm0.01$   \\
0411662605 & 43 & Lup & 244.4027 & -50.3948 &  10.6 &  $1.31 \pm 0.08$ &  $1.13 \pm 0.02$ & $0.50\pm0.05$   & $\mathcal{U}(1.4,   1.6)$ &   $\mathcal{U}(1631.0, 1632.0)$ &  $0.42\pm0.10$   & $0.36\pm0.10$ &    $0.07\pm0.01$   \\
\hline
\end{tabular}

  }
  \label{tab:2}
\end{table*}
\begin{figure*}
  \includegraphics[bb=20 161 577 718, width=0.9\textwidth]{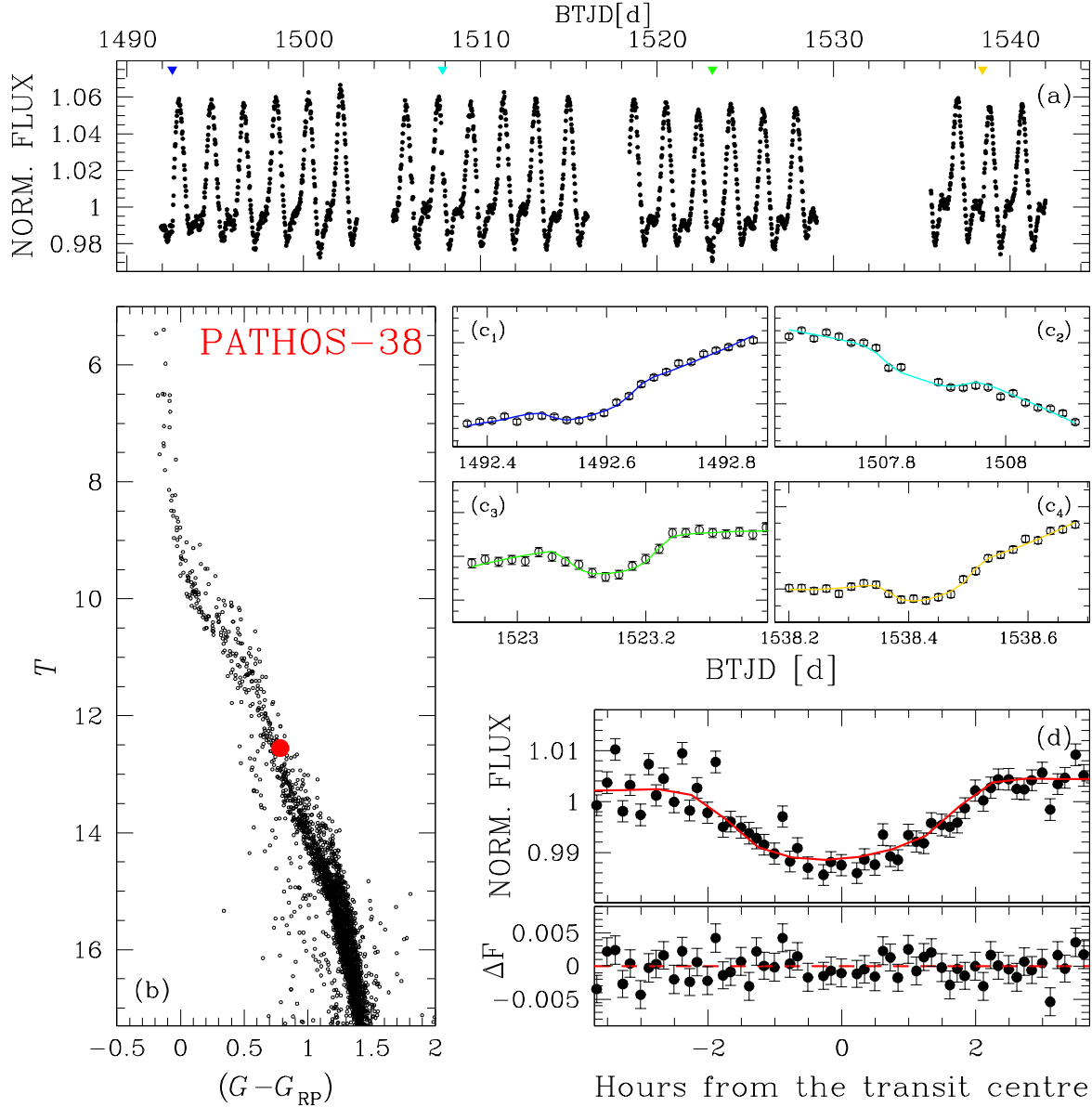}
  \caption{ Overview on the modelling procedure adopted for deriving
    the physical parameters of the transiting object PATHOS-38. Panel (a) shows
    the light curve of PATHOS-38 (TIC~123755508), whose position in
    the $T$ versus $G-G_{\rm RP}$ CMD is shown with a red circle in
    panel (b). Panels (c) show the local polynomial fit $+$ transit
    fit for each transit of the candidate exoplanet observed by {\it
      TESS}: the models are colour-coded as the arrows that indicate
    the epoch of the transit centres in panel (a). Panel (d) shows the
    folded transits (black points) after removing the local polynomial
    fits: in red the mean model of the transits. Bottom panel shows
    the difference between the observed and the modelled
    transits (see text for details). \label{fig:7}}
\end{figure*}

\begin{table*}
  \renewcommand{\arraystretch}{1.5}
  \caption{Results of transit modelling}
  \label{tab:3}
    \resizebox{.99\textwidth}{!}{
      \begin{tabular}{l c l r r r r r r r r r l}
  \hline
  TIC & PATHOS & Assoc. & \multicolumn{1}{c}{$P$} & \multicolumn{1}{c}{$T_0$}  & \multicolumn{1}{c}{$R_{\rm p}/R_{\star}$} & \multicolumn{1}{c}{$b$} &  \multicolumn{1}{c}{$a$}  &  \multicolumn{1}{c}{$\rho_{\star}$}   &  \multicolumn{1}{c}{$i$}   &  \multicolumn{1}{c}{$R_{\rm p}$}  &  \multicolumn{1}{c}{$R_{\rm p}$} &    Note  \\
      &        &        & \multicolumn{1}{c}{(d)} & \multicolumn{1}{c}{(BTJD)} &                                         &                         &  \multicolumn{1}{c}{(au)} & \multicolumn{1}{c}{($\rho_{\sun}$)}  & \multicolumn{1}{c}{(deg)}  & \multicolumn{1}{c}{($R_{\rm J}$)} & \multicolumn{1}{c}{($R_{\earth}$)} &   \\
  \hline
0081353413 & 35 & Vel & $  2.0865^{-0.0004}_{+0.0004}$  &   $1492.869^{-0.004}_{+ 0.003}$  & $0.132^{-0.0069}_{+ 0.009}$ & $0.31^{ -0.21}_{+  0.33}$ &  $0.0286^{-0.0007}_{+0.0006}$ &  $0.51^{ -0.23}_{+  0.13}$  & $86.8^{  -4.9}_{+   2.2}$ &  $1.42^{ -0.13}_{+  0.42} $  &  $15.9^{  -1.4}_{+   4.7}  $ &        \\
0081419525 & 36 & Vel & $  1.5498^{-0.0003}_{+0.0003}$  &   $1518.383^{-0.003}_{+ 0.003}$  & $0.053^{-0.0017}_{+ 0.002}$ & $0.37^{ -0.16}_{+  0.11}$ &  $0.0348^{-0.0002}_{+0.0002}$ &  $0.50^{ -0.01}_{+  0.01}$  & $85.3^{  -1.4}_{+   2.1}$ &  $0.87^{ -0.03}_{+  0.03} $  &  $  9.7^{  -0.3}_{+   0.3} $ &   (1)  \\
0095003423 & 37 & Lup & $  1.5268^{-0.0003}_{+0.0003}$  &   $1630.461^{-0.003}_{+ 0.003}$  & $0.202^{-0.0094}_{+ 0.011}$ & $0.51^{ -0.18}_{+  0.13}$ &  $0.0238^{-0.0004}_{+0.0004}$ &  $0.58^{ -0.12}_{+  0.11}$  & $83.7^{  -2.2}_{+   2.4}$ &  $2.16^{ -0.20}_{+  0.28} $  &  $ 24.2^{  -2.2}_{+   3.1} $ &        \\
0123755508 & 38 & Vel & $ 15.2869^{-0.0020}_{+0.0020}$  &   $1492.575^{-0.004}_{+ 0.004}$  & $0.128^{-0.0059}_{+ 0.007}$ & $0.77^{ -0.05}_{+  0.05}$ &  $0.1256^{-0.0011}_{+0.0011}$ &  $0.52^{ -0.09}_{+  0.09}$  & $87.9^{  -0.3}_{+   0.2}$ &  $1.61^{ -0.13}_{+  0.19} $  &  $ 18.1^{  -1.5}_{+   2.1} $ &        \\
0238235254 & 39 & Vel & $  5.5705^{-0.0004}_{+0.0004}$  &   $1497.200^{-0.003}_{+ 0.003}$  & $0.127^{-0.0166}_{+ 0.015}$ & $0.97^{ -0.02}_{+  0.02}$ &  $0.1243^{-0.0009}_{+0.0008}$ &  $0.14^{ -0.01}_{+  0.01}$  & $81.9^{  -0.3}_{+   0.3}$ &  $4.78^{ -0.65}_{+  0.59} $  &  $ 53.6^{  -7.3}_{+   6.6} $ &   (2)  \\
0238379370 & 40 & Vel & $ 27.1769^{-0.0072}_{+0.0067}$  &   $1511.748^{-0.009}_{+ 0.009}$  & $0.079^{-0.0089}_{+ 0.018}$ & $0.89^{ -0.05}_{+  0.05}$ &  $0.1895^{-0.0026}_{+0.0025}$ &  $0.49^{ -0.07}_{+  0.07}$  & $88.3^{  -0.2}_{+   0.1}$ &  $1.05^{ -0.14}_{+  0.27} $  &  $ 11.8^{  -1.5}_{+   3.0} $ &        \\
0307610438 & 41 & Lup & $  1.5615^{-0.0003}_{+0.0003}$  &   $1601.955^{-0.002}_{+ 0.002}$  & $0.072^{-0.0026}_{+ 0.003}$ & $0.80^{ -0.04}_{+  0.04}$ &  $0.0305^{-0.0008}_{+0.0008}$ &  $0.20^{ -0.03}_{+  0.03}$  & $76.1^{  -1.5}_{+   1.3}$ &  $1.38^{ -0.10}_{+  0.13} $  &  $ 15.5^{  -1.2}_{+   1.4} $ &   (1)  \\
0374732772 & 42 & Lup & $ 18.7944^{-0.0092}_{+0.0042}$  &   $1632.117^{-0.012}_{+ 0.015}$  & $0.077^{-0.0120}_{+ 0.018}$ & $0.77^{ -0.21}_{+  0.13}$ &  $0.1494^{-0.0016}_{+0.0016}$ &  $0.44^{ -0.05}_{+  0.05}$  & $88.0^{  -0.4}_{+   0.5}$ &  $1.06^{ -0.17}_{+  0.26} $  &  $ 11.9^{  -1.9}_{+   3.0} $ &   (3)  \\
0411662605 & 43 & Lup & $  1.5610^{-0.0006}_{+0.0006}$  &   $1631.386^{-0.004}_{+ 0.005}$  & $0.056^{-0.0023}_{+ 0.002}$ & $0.17^{ -0.12}_{+  0.16}$ &  $0.0274^{-0.0002}_{+0.0002}$ &  $0.47^{ -0.05}_{+  0.05}$  & $87.8^{  -2.1}_{+   1.5}$ &  $0.73^{ -0.04}_{+  0.04} $  &  $  8.2^{  -0.4}_{+   0.4} $ &   (4)     \\
\hline
\multicolumn{13}{l}{$^{(1)}$~Also in the TOI catalogue} \\
\multicolumn{13}{l}{$^{(2)}$~Radius too large, suspected eclipsing binary} \\
\multicolumn{13}{l}{$^{(3)}$~Another single deeper (suspected) transit ($\delta_{\rm T}\sim 5$~\%) is present in the light curve due to a likely stellar companion or second planet. } \\
\multicolumn{13}{l}{$^{(4)}$~Likely field star. } \\
\end{tabular}

    }
\end{table*}

\subsection{Modelling of the transits}

I used the python package
\texttt{PYORBIT}\footnote{\url{https://github.com/LucaMalavolta/PyORBIT}}
(\citealt{2016A&A...588A.118M,2018AJ....155..107M,2019A&A...630A..81B}),
developed for the modelling of planetary transits and radial
velocities. The routine is based on the combined use of the package \texttt{BATMAN} (\citealt{2015PASP..127.1161K}), the affine invariant Markov chain Monte
Carlo (MCMC) sampler \texttt{EMCEE} (\citealt{2013PASP..125..306F}), and
the global optimization algorithm \texttt{PYDE}\footnote{\url{https://github.com/hpparvi/PyDE}}(\citealt{1997..............S}).

For the transit model, I included the central time of the first
transit ($T_0$), the period ($P$), the impact parameter ($b$), the
planetary-to-stellar-radius ratio ($R_{\rm P}/R_{\star}$), the stellar
density ($\rho_{\star}$). To locally model the stellar activity, for each
transit a 2nd-degree polynomial fit is performed to the out-of-transit
part of the light curve. To take into account the impact of the
dilution ($df$) on the error estimate of $R_{\rm P}$, I included in the
modelling this quantity as a free parameter, with a Gaussian prior
obtained by using the stars in the Gaia~DR2 catalogue that fall in the
same pixel of the target, and transforming their Gaia magnitude in
{\it TESS} magnitudes with the equations reported by
\citet{2019AJ....158..138S}. I extracted information on the limb
darkening (LD) coefficients by using the $T_{\rm eff}$ and $\log(g)$
values obtained with the isochrone fitting, and the grid of values
published by \citet{2018A&A...618A..20C}; the LD parametrization
adopted is that described by \citet{2013MNRAS.435.2152K}.  All the
priors adopted for the modelling are listed in Table~\ref{tab:2}.  For the modelling of the transits I adopted a circular orbit ($e=0$). In
the modelling process, the routine took into account of the 30-min
cadence of the TESS time-series (\citealt{2010MNRAS.408.1758K}). The
routine explored all the parameters in linear space, using a number of
walkers $N_{\rm walkers}$ equal to 10 times the number of free
parameters. For each model, I run the sampler for 80\,000 steps,
cutting away the first 15\,000 steps as burn-in, and using a thinning
factor of 100.

\begin{figure*}
  \includegraphics[bb=22 179 570 701, width=0.7\textwidth]{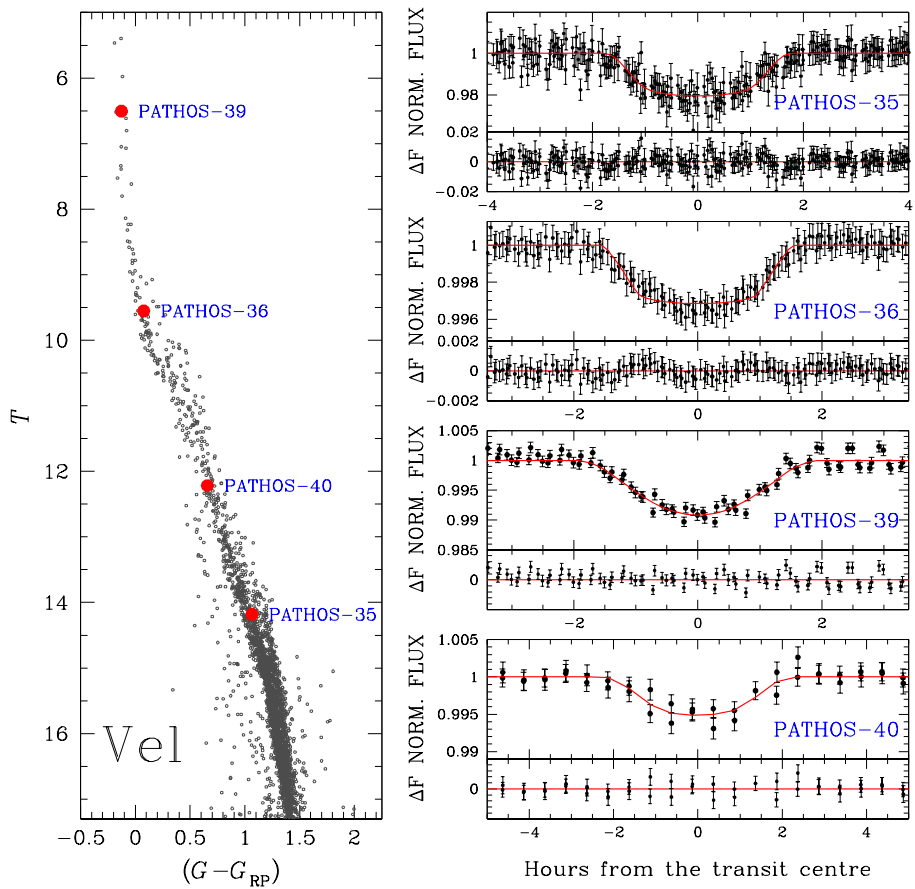}
  \caption{Overview on the transiting objects (PATHOS-35, 36, 39, and
    40) orbiting Vel association stars. Left panel shows the $G$
    versus $(G-G_{\rm RP})$ CMD of the likely Vel members: red dots
    indicate the transiting objects' positions. Right panels show
    the phased transits of the objects of interest and the fitted transit
    models (red); for each object, the difference between the
    observations and the model is shown below its folded light curve.
    \label{fig:8a}}
\end{figure*}

\begin{figure*}
  \includegraphics[bb=22 179 570 701, width=0.7\textwidth]{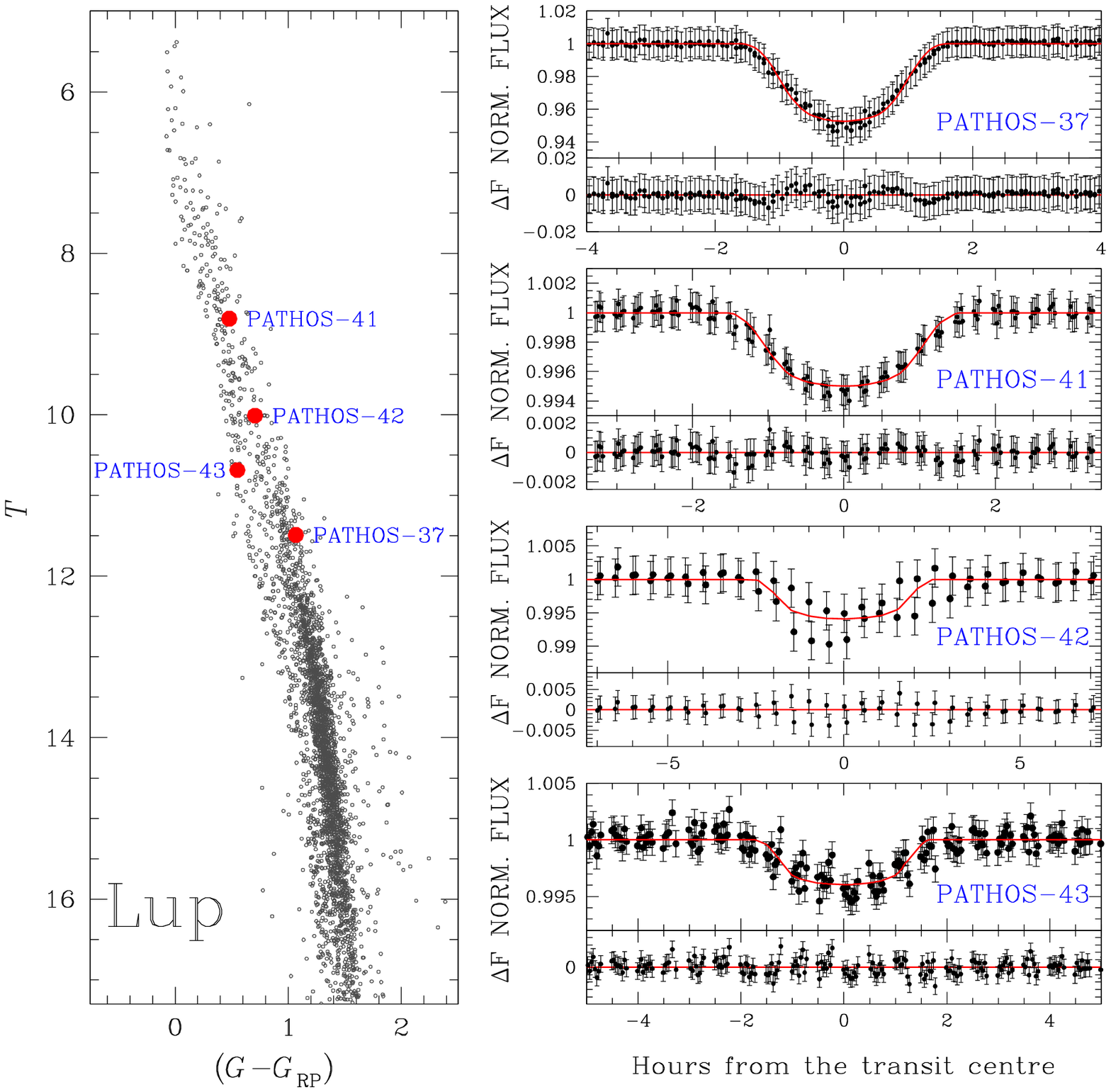}
  \caption{As in Fig.~\ref{fig:8a}, but for objects of interest that are likely
    members of Lup association (PATHOS-37, 41, 42, 43).
    \label{fig:8b}}
\end{figure*}

An overview on the modelling procedure is reported in
Fig.~\ref{fig:7} for PATHOS-38. The results of the modelling for all 
the objects of interest are reported in Table~\ref{tab:3} and Figs.~\ref{fig:8a} and \ref{fig:8b}.



\subsection{Candidate exoplanets' frequency in young associations}
In this work I found and modelled 9 transiting objects of
interest. For the analysis described in this section, I excluded the
objects with a radius $R_{\rm P}\gtrsim 2~R_{\rm J}$ (PATHOS-37 and
PATHOS-39), because of their doubtful planet nature. I excluded also
PATHOS-43 because, on the basis of its position on the $G$ versus
$(G-G_{\rm RP})$ CMD (see Fig.~\ref{fig:8b}), it has a high
probability to be not an association member.

All the 6 survived candidates I have detected are Jupiter size
candidate exoplanets ($R_{\rm P} \sim 0.9$--$1.6~R_{\rm J}$). No
Neptune- or Earth-size candidate planets have been detected. Given the
distance of the associations ($\sim 150$--$400$~pc) and the
photometric precision of the light curves, on the basis of the analysis performed in
\citetalias{2020MNRAS.495.4924N} (see Fig.~8), it is not possible to
detect (super-)Earth size planets around association members studied in this work. On
the basis of the same analysis, it is possible to detect Neptune-size
exoplanets only for stars with radii $R_{\star} \lesssim
1.0~R_{\sun}$. I calculated the expected number of exoplanets ($N_{\rm
  planet}$) as done in \citetalias{2020MNRAS.495.4924N}, by using the
(modified) equation:
\begin{equation}
  N_{\rm planet} = f_{\star} \times \sum_r{{N_{\star}}^r \times {{\rm Pr}_{\rm transit}}^r}
\label{eq:2}
\end{equation}
where $f_{\star}$ is the percentage of stars with at least one
exoplanet, the sum on $r$ indicate the intervals of stellar radii
considered
($r=[0.0,0.5],[0.5,1.0],[1.0,1.5],[1.5,2.0]~R_{\sun}$),
$N_{\star}$ is the number of stars in the considered stellar radius
bins, ${\rm Pr}_{\rm transit}\simeq R_{\star}/a$ is the transit
probability, with $a$ the semi-major axis of the orbit, calculated
using the third law of Kepler and by using an average period
$P=10$~d. Considering: (1) the stars for which I analysed the light
curves; (2) the stars with $R_{\star} \le 1 R_{\sun}$; (3) the
frequencies $f_{\star}$ for Neptune-size exoplanets tabulated by
\citet{2013ApJ...766...81F} in the case of exoplanets with
$P=0.8-10.0$~d ($\sim 0.21$~\%) and calculated in
\citetalias{2020MNRAS.495.4924N} ($\sim 1.34$~\%), I expect to find
$N_{\rm planet}(R_{\rm P}=1~R_{\rm N})= 1 \pm 1$, in agreement with
the null detection found in this work.

The Jupiter size candidates found in this work orbit stars in the Lup
(2 candidates) and Vel (4 candidates) associations, the oldest
associations studied in this work, while no candidates have been found
around ChI, ChII, and CrA associations.  By using
equation~(\ref{eq:2}), I calculated the frequency of candidate Jupiters
in Lup and Vel associations. Because the periods of the candidates
range between $\sim 1.5$~d and $\sim 27$~d, I divided the sample of
candidate exoplanets in three sub-samples and, on the basis of their
periods, I calculated the frequency $f_{\star}$ by using the
transit probabilities associated to the mean period $\bar{P}$ of the candidates that form the sub-sample:

\noindent
(i) for candidates with period $1.0~{\rm d}\le P \le 2.1~{\rm d}$,
($\bar{P}\sim 1.7$\,d) I found $f_{\star}=(0.70 \pm 0.70)$~\% and
$f_{\star}=(0.44 \pm 0.31)$~\%, for Lup and Vel association,
respectively; considering all the stars analysed in this work (i.e.,
including also the ChI, ChII, and CrA members), I found
$f_{\star}=(0.43 \pm 0.25)$~\%. For giant planets orbiting field stars
with periods $0.8~{\rm d}\le P \le 2.0~{\rm d}$,
\citet{2013ApJ...766...81F} tabulated a frequency of
$f_{\star}=(0.015\pm 0.007)$~\%, that is lower than the mean values
found in this work but in agreement within $\sim 2\sigma$;

\noindent
(ii) for candidates with period $15.0~{\rm d}\le P \le 30~{\rm d}$
($\bar{P} \sim 20.3$~d) I found $f_{\star}=(3.64 \pm 3.64)$~\%,
$f_{\star}=(2.28 \pm 1.63)$~\%, and $f_{\star}=(2.24 \pm 1.31)$~\% if
I consider only Lup members, Vel members, and all the stars,
respectively. For Jupiter exoplanets orbiting field stars with period
in the range $17.0~{\rm d}\le P \le 29.0~{\rm d}$,
\citet{2013ApJ...766...81F} found a frequency $f_{\star}=(0.23 \pm
0.12)$~\%; also in this case the mean frequency found in this work is
higher than that tabulated by \citet{2013ApJ...766...81F}, even if
they agree within $2\sigma$.

\noindent
(iii) for candidates with period $1.0~{\rm d}\le P \le 30~{\rm d}$
($\bar{P} \sim 11.0$~d) the frequencies of Jupiter-size exoplanets in
the Lup and Vel members are $f_{\star}=(4.84 \pm 3.51)$~\% and
$f_{\star}=(3.03 \pm 1.54)$~\%, respectively. Considering all the
stars analysed in this work, I found a frequency $f_{\star}=(2.99 \pm
1.24)$~\%. \citet{2013ApJ...766...81F} found for giant planets around
field stars with periods $0.8~{\rm d}\le P \le 29.0~{\rm d}$ a
frequency $f_{\star}=(0.93 \pm 0.10)$~\%, also in this case lower than
the value found in this work, but in agreement within $2 \sigma$. \\
I want to emphasise that, for the statistical analysis performed in this
work, the completeness of the detection method was not taken into
account, and therefore the calculated frequencies might be considered
as lower limits.

By using the results previously obtained, I calculated how many
transiting Jupiters are expected to be found in the others three
associations: even considering the maximum mean frequency found in the
previous analysis ($\sim 4.8$~\%), the expected number of giants in
the three associations is $N_{\rm P}<1$, in agreement with the null
detection obtained in this work.

\section{Summary and Conclusion}
\label{sec:sum}

In the present work, the third of the PATHOS project, I performed a
detailed analysis of the light curves of stars in five young
({\it T-})associations associated to star forming regions:
Chamaeleon I and II, Lupus, Corona Australis, and $\gamma$ Velorum
association. These associations have been chosen because of their
young age ($\lesssim 10$~Myr): indeed, searching and characterising
exoplanets orbiting very young stars allow us to constraint
theoretical models on the formation of them and to understand the
mechanisms that prevail in their dynamical and physical evolution
(migration, atmospheric loss, etc.).

For this work, I extracted and corrected 7150 light curves of 4459
association members from {\it TESS} FFIs by using the PSF-based
approach pipeline already adopted with success in previous
works. Light curves will be publicly available as HLSP on the PATHOS
project
webpage\footnote{\url{https://archive.stsci.edu/hlsp/pathos}}(DOI:
10.17909/t9-es7m-vw14) of the MAST archive.

By performing an analysis of the GLS periodograms of the light curves,
I identified 1260 periodic variable stars. Combining the
gyrochronological analysis of periodic variable stars and the
isochrone fitting of the CMD, I constrained the ages of the
associations, obtaining that the five associations have ages that span
between $\sim 2$\,Myr and $\sim 10$\,Myr. Because of the young age,
the analysed associations host a large number of YSOs surrounded by
circumstellar disk. By the analysis of the light curves, I identified
71 dipper stars, i.e., stars that present in their light curves
important drops of the flux on timescales of $\lesssim 1$ day. These
drops of the luminosity are due to the dust that form the inner
regions of the circumstellar disks; comparing the simultaneous drops
observed in {\it TESS} and $g$-band ASAS-SN light curves, I calculated
the ratio between the absorptions in $T$ and $g$ bands ($A_T/A_g$),
that gives us information on the size of the grains that form the
disk. In particular, when $A_T/A_g \to 1 $ the grains have sizes
comparable with the wavelengths in which {\it TESS} observes; lower
values of the ratio $A_T/A_g$ are associated to smaller grains. I
found a weak anti-correlation between $A_T/A_g$ and the dereddened
colour $(V-K)_0$, with grain sizes that decrease with the mass of the
hosting star. This work can not confirm the correlation between the infrared
excess $(K_s-{\rm [22 \mu m]})$ and $A_T/A_g$ found by
\citet{2020MNRAS.tmp.1773B}. Finally, I found that the highest
frequency of dippers are associated to the low mass stars of the
youngest associations (ChI and ChII, $\sim 2.5$~Myr, frequency $\sim
20$\,\%), and that the frequency of dippers is anti-correlated with
the age of the associations, confirming that the timescales for the
disk cleaning around low-mass stars is $<10$\,Myr
(\citealt{2012ApJ...758...31L}).

I searched for transit signals among the light curves of the
association members, and after the vetting tests (analysis of the
odd/even transits, of the in-/out-of transit centroid offset, etc.), 9
objects of interest passed the selections. In order to derive the
physical parameters of the transiting objects, I modelled their
transits by using their light curves and the stellar parameters
derived through isochrone fits. Excluding two objects of interest, because
their radius is too large ($R_{\rm P}>2 R_J$), and another object of
interest because hosted by a likely field star, I detected 6 Jupiter
size candidates: 2 in the Lup association and 4 in the Vel
association. No Earth, super-Earth, and Neptune size candidates have
been detected; anyway, given the distance of the associations, the
number of members, and the frequency of these kind of exoplanets
tabulated by \citet{2013ApJ...766...81F} and in
\citetalias{2020MNRAS.495.4924N}, the null detection is agreement
with the expectations. The mean frequency of giant planets in
associations derived considering different period intervals ranges
between $\sim 1$~\% and $\sim 4$~\%, higher than the values reported by
\citet{2013ApJ...766...81F} for giants orbiting field stars ($\lesssim
1$\%) and in \citetalias{2020MNRAS.495.4924N} for Jupiters orbiting
open cluster members ($\sim 0.20$~\%). Anyway, given the low number of
candidates, the errors on the calculated frequencies are too large,
and the obtained results must be considered provisional. I also verified
 if the null detection of giant planets around ChI, ChII and CrA
members is expected: even considering a frequency 
$f_{\star}\sim 5$~\%, the number of Jupiter size exoplanet expected is
$N_{\rm P}<< 1$, in agreement with the null detection of this work.

The analysis of the light curves of older association members
($\sim$10--100\,Myr), targets of next works of the PATHOS project, is
mandatory to understand if the frequency and/or the orbital and
physical parameters of the exoplanets are correlated with the age of
the hosting stars. In this way, it will be possible understand how
exoplanets born, and trace the prevailing mechanisms that characterise
their life.

\appendix

\section{Electronic material}

\begin{table*}[t!]
  \caption{Description of the column content of the catalogue of variable stars.}
  \resizebox{0.75\textwidth}{!}{
    \begin{tabular}{ l c c l}
\hline
\multicolumn{1}{c}{Column} &
\multicolumn{1}{c}{Name} &
\multicolumn{1}{c}{Unit} &
\multicolumn{1}{c}{Explanation} \\
\hline
01  & \texttt{RA}         & [deg.]   &  Right ascension (J2000, epoch 2015.5)  \\
02  & \texttt{DEC}        & [deg.]   &  Declination (J2000, epoch 2015.5) \\
03  & \texttt{TIC}        &          &  {\it TESS} Input Catalogue ID \\
04  & \texttt{GAIA\_DR2}  &          &  Gaia DR2 Source ID \\
05  & \texttt{PERIOD}     & [d]      &  Period             \\
06  & \texttt{Gmag     }  & [mag]    &  Gaia DR2 $G$ magnitude\\
07  & \texttt{e\_Gmag   } & [mag]    &  Error on Gaia DR2 $G$ magnitude\\
08  & \texttt{BPmag    }  & [mag]    &  Gaia DR2 $G_{\rm BP}$ magnitude\\
09  & \texttt{e\_BPmag  } & [mag]    &  Error on Gaia DR2 $G_{\rm BP}$ magnitude\\
10  & \texttt{RPmag    }  & [mag]    &  Gaia DR2 $G_{\rm RP}$ magnitude\\
11  & \texttt{e\_RPmag  } & [mag]    &  Error on Gaia DR2 $G_{\rm RP}$ magnitude\\
12  & \texttt{Tmag     }  & [mag]    &  {\it TESS} $T$ magnitude\\
13  & \texttt{e\_Tmag   } & [mag]    &  Error on {\it TESS} $T$ magnitude\\
14  & \texttt{Bmag     }  & [mag]    &  $B$-Johnson magnitude\\
15  & \texttt{e\_Bmag   } & [mag]    &  Error on $B$-Johnson magnitude\\
16  & \texttt{Vmag     }  & [mag]    &  $V$-Johnson magnitude\\
17  & \texttt{e\_Vmag   } & [mag]    &  Error on $V$-Johnson magnitude\\
18  & \texttt{Jmag     }  & [mag]    &  2MASS $J$ magnitude\\
19  & \texttt{e\_Jmag   } & [mag]    &  Error on 2MASS $J$ magnitude\\
20  & \texttt{Hmag     }  & [mag]    &  2MASS $H$ magnitude\\
21  & \texttt{e\_Hmag   } & [mag]    &  Error on 2MASS $H$ magnitude\\
22  & \texttt{Kmag     }  & [mag]    &  2MASS $K_s$ magnitude\\
23  & \texttt{e\_Kmag   } & [mag]    &  Error on 2MASS $K_s$ magnitude\\
24  & \texttt{W1mag    }  & [mag]    &  WISE $W_1$ magnitude\\
25  & \texttt{e\_W1mag  } & [mag]    &  Error on WISE $W_1$ magnitude\\
26  & \texttt{W2mag    }  & [mag]    &  WISE $W_2$ magnitude\\
27  & \texttt{e\_W2mag  } & [mag]    &  Error on WISE $W_2$ magnitude\\
28  & \texttt{W3mag    }  & [mag]    &  WISE $W_3$ magnitude\\
29  & \texttt{e\_W3mag  } & [mag]    &  Error on WISE $W_3$ magnitude\\
30  & \texttt{W4mag    }  & [mag]    &  WISE $W_4$ magnitude\\
31  & \texttt{e\_W4mag  } & [mag]    &  Error on WISE $W_4$ magnitude\\
32  & \texttt{E\_BV }     &          &  $E(B-V)$ \\
33  & \texttt{PARALLAX}   & mas      &  Parallax from Gaia~DR2 \\
34  & \texttt{PM\_RA}     & mas~yr$^{-1}$   &  Proper motion along the RA direction from Gaia~DR2 \\
35  & \texttt{PM\_DEC}    & mas~yr$^{-1}$   &  Proper motion along the DEC direction from Gaia~DR2 \\
36  & \texttt{ASSOCIATION} &          &  Name of the association that host the star \\
\hline
\end{tabular}

  }
  \label{tab:a1}
\end{table*}

The catalogues of the periodic variable stars and of the dippers
analysed in this work are available electronically as supporting
material to this paper. Both the catalogues are in \texttt{ascii} and
\texttt{fits} format. A description of the columns for the two
catalogues are reported in Tables~\ref{tab:a1} and ~\ref{tab:a2}.

Light curves extracted and analysed in this work are available in the
MAST archive as HLSP under the project
PATHOS\footnote{\url{https://archive.stsci.edu/hlsp/pathos}} (DOI:
10.17909/t9-es7m-vw14). The updated list of candidate exoplanets is
reported on the PATHOS webpage of the MAST archive.

\begin{table*}
  \caption{Description of the column content of the catalogue of dipper stars.}
  \resizebox{0.75\textwidth}{!}{
  \begin{tabular}{ l c c l}
\hline
\multicolumn{1}{c}{Column} &
\multicolumn{1}{c}{Name} &
\multicolumn{1}{c}{Unit} &
\multicolumn{1}{c}{Explanation} \\
\hline
01  & \texttt{RA}         & [deg.]   &  Right ascension (J2000, epoch 2015.5)  \\
02  & \texttt{DEC}        & [deg.]   &  Declination (J2000, epoch 2015.5) \\
03  & \texttt{TIC}        &          &  {\it TESS} Input Catalogue ID \\
04  & \texttt{GAIA\_DR2}  &          &  Gaia DR2 Source ID \\
05  & \texttt{AT\_AG}     &          &  $A_{T}/A_{g}$ values [$-99.9$: not available] \\
06  & \texttt{e\_ATAG}    &          &  Error on $A_{T}/A_{g}$ values [$-99.9$: not available] \\
07  & \texttt{Gmag     }  & [mag]    &  Gaia DR2 $G$ magnitude\\
08  & \texttt{e\_Gmag   } & [mag]    &  Error on Gaia DR2 $G$ magnitude\\
09  & \texttt{BPmag    }  & [mag]    &  Gaia DR2 $G_{\rm BP}$ magnitude\\
10  & \texttt{e\_BPmag  } & [mag]    &  Error on Gaia DR2 $G_{\rm BP}$ magnitude\\
11  & \texttt{RPmag    }  & [mag]    &  Gaia DR2 $G_{\rm RP}$ magnitude\\
12  & \texttt{e\_RPmag  } & [mag]    &  Error on Gaia DR2 $G_{\rm RP}$ magnitude\\
13  & \texttt{Tmag     }  & [mag]    &  {\it TESS} $T$ magnitude\\
14  & \texttt{e\_Tmag   } & [mag]    &  Error on {\it TESS} $T$ magnitude\\
15  & \texttt{Bmag     }  & [mag]    &  $B$-Johnson magnitude\\
16  & \texttt{e\_Bmag   } & [mag]    &  Error on $B$-Johnson magnitude\\
17  & \texttt{Vmag     }  & [mag]    &  $V$-Johnson magnitude\\
18  & \texttt{e\_Vmag   } & [mag]    &  Error on $V$-Johnson magnitude\\
19  & \texttt{Jmag     }  & [mag]    &  2MASS $J$ magnitude\\
20  & \texttt{e\_Jmag   } & [mag]    &  Error on 2MASS $J$ magnitude\\
21  & \texttt{Hmag     }  & [mag]    &  2MASS $H$ magnitude\\
22  & \texttt{e\_Hmag   } & [mag]    &  Error on 2MASS $H$ magnitude\\
23  & \texttt{Kmag     }  & [mag]    &  2MASS $K_s$ magnitude\\
24  & \texttt{e\_Kmag   } & [mag]    &  Error on 2MASS $K_s$ magnitude\\
25  & \texttt{W1mag    }  & [mag]    &  WISE $W_1$ magnitude\\
26  & \texttt{e\_W1mag  } & [mag]    &  Error on WISE $W_1$ magnitude\\
27  & \texttt{W2mag    }  & [mag]    &  WISE $W_2$ magnitude\\
28  & \texttt{e\_W2mag  } & [mag]    &  Error on WISE $W_2$ magnitude\\
29  & \texttt{W3mag    }  & [mag]    &  WISE $W_3$ magnitude\\
30  & \texttt{e\_W3mag  } & [mag]    &  Error on WISE $W_3$ magnitude\\
31  & \texttt{W4mag    }  & [mag]    &  WISE $W_4$ magnitude\\
32  & \texttt{e\_W4mag  } & [mag]    &  Error on WISE $W_4$ magnitude\\
33  & \texttt{E\_BV }     &          &  $E(B-V)$ \\
34  & \texttt{PARALLAX}   & mas      &  Parallax from Gaia~DR2 \\
35  & \texttt{PM\_RA}     & mas~yr$^{-1}$   &  Proper motion along the RA direction from Gaia~DR2 \\
36  & \texttt{PM\_DEC}    & mas~yr$^{-1}$   &  Proper motion along the DEC direction from Gaia~DR2 \\
37  & \texttt{ASSOCIATION} &          &  Name of the association that host the star \\
\hline
\end{tabular}

  }
  \label{tab:a2}
\end{table*}

\section*{Acknowledgements}
I acknowledge the support from the French Centre National d'Etudes
Spatiales (CNES). I acknowledge the partial support from PLATO
ASI-INAF agreements n. 2015-019-R0-2015 and 2015-019-R.1-2018.
I warmly thank the referee for carefully reading the manuscript.
I thank M.~Deleuil and G.~Piotto for the useful suggestions on
this work, and L.~Malavolta for his support in the use of \texttt{PYORBIT}.
This paper includes data collected by the {\it TESS} mission. Funding for the
{\it TESS} mission is provided by the NASA Explorer Program. This work has
made use of data from the European Space Agency (ESA) mission {\it
  Gaia} (\url{https://www.cosmos.esa.int/gaia}), processed by the {\it
  Gaia} Data Processing and Analysis Consortium (DPAC,
\url{https://www.cosmos.esa.int/web/gaia/dpac/consortium}). Funding
for the DPAC has been provided by national institutions, in particular
the institutions participating in the {\it Gaia} Multilateral
Agreement. Some tasks of the data analysis have been carried out using
\texttt{VARTOOLS} v~1.38 (\citealt{2016A&C....17....1H}) and
\texttt{TLS} python routine (\citealt{2019A&A...623A..39H}).

\section*{Data Availability}
The data underlying this article are available in MAST at
\url{doi:10.17909/t9-es7m-vw14} and at
\url{https://archive.stsci.edu/hlsp/pathos}. \\ The data underlying
this article are available in the article and in its online
supplementary material.


\bibliographystyle{mnras}
\bibliography{biblio}


\label{lastpage}
\end{document}